\documentclass[journal,article,submit,pdftex,moreauthors]{Definitions/mdpi}

\firstpage{1} 
\pubvolume{1}
\issuenum{1}
\articlenumber{0}
\pubyear{2023}
\copyrightyear{2023}
\datereceived{ } 
\daterevised{ } 
\dateaccepted{ } 
\datepublished{ } 
\hreflink{https://doi.org/} 

\Title{Decoding the nature of Coherent radio emission in Pulsars I: Observational constraints}

\TitleCitation{Title}

\Author{Dipanjan Mitra $^{1,2,*}$\orcidA{}, Rahul Basu $^{2}$\orcidB{} and George I Melikizde $^{2,3}$\orcidC{}}

\AuthorNames{Firstname Lastname, Firstname Lastname and Firstname Lastname}

\address{%
$^{1}$ \quad National Centre for Radio Astrophysics, Tata Institute of Fundamental Research, Pune 411007, India.\\
$^{2}$ \quad Janusz Gil Institute of Astronomy, University of Zielona G\'ora, ul. Szafrana 2, 65-516 Zielona G\'ora, Poland.\\
$^{3}$ \quad Evgeni Kharadze Georgian National Astrophysical Observatory, 0301 Abastumani, Georgia.}

\corres{Correspondence: dmitra@ncra.tifr.res.in}

\abstract{Radio observations from normal pulsars indicate that the
  coherent radio emission is excited by curvature radiation from
  charge bunches. In this review we provide a systematic description
  of the various observational constraints on the radio emission
  mechanism. We have discussed the presence of highly polarized time
  samples where the polarization position angle follow two orthogonal
  well defined tracks across the profile, that closely match the
  rotating vector model in an identical manner. The observations also
  show the presence of circular polarization, with both the right and
  left handed circular polarization seen across the profile. Other
  constraints on the emission mechanism is provided by the detailed
  measurements of the spectral index variation across the profile
  window, where the central part of the profile, corresponding to the
  core component, has a steeper spectrum than the surrounding cones.
  Finally, the detailed measurements of the subpulse drifting
  behaviour can be explained by considering the presence of
  non-dipolar field on the stellar surface and the formation of the
  Partially screened Gap (PSG) above the polar cap region. The PSG
  gives rise to a non-stationary plasma flow, that has a
  multi-component nature, consisting of highly energetic primary
  particles, secondary pair plasma and iron ions discharged from the
  surface, with large fragmentation resulting is dense plasma clouds
  and lower density inter-cloud regions. The physical properties of
  the outflowing plasma and the observational constraints lead us to
  consider coherent curvature radiation as the most viable explanation
  for the emission mechanism in normal pulsars, where propagation
  effects due to adiabatic walking and refraction are largely
  inconsequential.}  

\begin{document}

\section{Introduction}

The discovery of pulsars \citep{1968Natur.217..709H} and their
association with rotating neutron stars
\cite{1969Natur.221...25G,1968Natur.220..753L,1968Sci...162.1481S,
  1969Natur.221..453C,1969Natur.221..525C,1969ApJ...155L.121L} raised
questions about the enormously high brightness temperatures ($\sim
10^{30}$ K) of the radio emission from these sources. These high
values exceed all limits of incoherent emission processes and require
the presence of a collective or coherent emission mechanism in pulsar
\cite{1969Ap&SS...4..464G}. Pulsar radio emission is highly polarized
and broadband in nature, usually measured between tens of MHz and tens
of GHz. It exhibits a wide range of time dependent phenomena like
subpulse drifting, emission mode changing, nulling, microstructures in
the single pulses, etc., that pose additional challenges for any
proposed model of the coherent emission mechanism.

The pulsar magnetosphere is filled with dense plasma composed of
electrons and positrons, and can be divided into open and closed
dipolar magnetic field line regions. An outflowing plasma with
relativistic energies is expected to stream along the open field line
regions where the radio emission is generated
\cite{1969ApJ...157..869G,1971ApJ...164..529S}.  Historically,
  two main classes of coherent radio mechanisms have been
  distinguished in pulsars: the maser and the antenna mechanisms
  \citep{1975ARA&A..13..511G,1975ApJ...196...51R}. The maser mechanism
  requires the development of some plasma instability due to specific
  distribution function, with likely candidates being the anomalous
  Doppler effect, the Cherenkov-drift instability, the free electron
  maser, the inverse Compton scattering, the linear acceleration, etc
  (\cite{1988VA.....31..393Q,1988hea..conf...88Q,
    2000ApJ...544.1081M,1991MNRAS.253..377K,
    1992RSPTA.341..105M,1995JApA...16..137M, 2021MNRAS.500.4530M}).
  The antenna mechanism is conventionally associated with coherent
  curvature radiation (CCR hereafter).  Such division however appear
  to be quite contrived, since any mechanism has to be associated with
  some kind of instability in the plasma that is capable of exciting
  plasma waves.  The resulting waves should then escape directly or
  through other processes from the magnetosphere (see
  \cite{1992RSPTA.341..105M}).  Despite advances made by these
  different models the exact process of how the coherent radio
  emission mechanism operates in pulsar magnetosphere remains hitherto
  an unsolved and challenging problem.  The requirements from
  observations that can delineate a proper mechanism for pulsar radio
  emission will be discussed in this article.

The eigen-modes excited in a plasma by any emission mechanism
propagate in the medium and eventually detach to reach the observer.
The properties of the eigen-modes in strongly magnetized, homogeneous,
pair plasma has been relatively well established (see e.g.
\cite{1978ApJ...219..274H, 1986ApJ...302..120A}), and briefly
summarized from the perspective of pulsars as follows. If the
distribution functions of electrons and positrons, the primary
constituents of the pulsar plasma, are identical, the dispersion
equations describing the system yield two orthogonally polarized modes
in the magnetized pair plasma, the ordinary (O-mode) and extraordinary
(X-mode) waves.  In this work we follow the nomenclature of
\cite{2003PhRvE..67b6407S} to describe the eigen-modes in the plasma.
The polarization vector of O-mode lies in the plane of the propagation
vector, $\vec{k}$, and the ambient magnetic field, $\vec{B}$, with a
component along $\vec{B}$ as well as $\vec{k}$. The X-mode, also known
as transverse, $t$-mode, has a purely non-potential nature with the
polarization vector directed perpendicular to the plane containing
$\vec{k}$ and $\vec{B}$. There are two additional branches of the
O-mode, $lt_{1}$-mode and $lt_{2}$-mode, that are mixed
longitudinal-transverse in nature. If we consider parallel
propagation, such that $\vec{k} \parallel \vec{B}$, the $lt_{2}$-mode
coincides with the Langmuir mode and has a purely potential character.
Two limiting frequencies of the Langmuir mode can be found in the
observer's frame, when $k=0$, we have $\omega_1 =
\omega_p/\gamma^{1.5}$ while in case of $k = \omega/c$ one obtains
$\omega_{\circ} = 2\sqrt{\gamma}~\omega_p$, here $\gamma$ is the
Lorentz factor, $\omega_p = \sqrt{2\pi n_p e^2/m_e}$ is the plasma
frequency, $n_p$ is the plasma density, $e$ is the electron charge and
$m_e$ is the mass of electrons. Both the $lt_{1}$ branch of the O-mode
and the X-mode have purely transverse characteristics with the
polarization vector either parallel or perpendicular to the
$\vec{k}\times\vec{B}$ plane. The phase velocity of $lt_{1}$-mode is
always sub-luminal, while $lt_{2}$-mode is super-luminal for
relatively small values of the wave vectors and can be sub-luminal at
higher frequencies. The $t$-modes are the most suitable candidate
  for the observed emission, since in presence of strong magnetic
  fields in the inner magnetosphere where the pulsar radio emission is
  generated, the $t$-modes can escape without any limitations (vacuum
  like).  However for certain excitation mechanisms, the $t$-modes
  encounters difficulties in explaining the observed linear
  polarization position angle behaviour across the pulse profile,
  where the position angle closely mimic the change in the magnetic
  field line planes of a rotating dipole magnet.  The difficulty
  arises because the polarization vector of the $t$-mode is
  perpendicular to the $\vec{k}$ and $\vec{B}$ planes, and if the
  exciting mechanism excites a range of $k-B$ planes, then the
  resultant position angle will not follow the change in the magnetic
  field line planes.  A defining attribute of the CCR mechanism is
  that the waves excited by this mechanism are polarised either
  perpendicular to the curved magnetic field plane or lies in this
  plane, and hence CCR can explain the observed position angle
  behaviour and hence is a strong candidate for pulsar emission
  mechanism.

There are a number of challenges in finding observational signatures
of the emission mechanism due to changes in both the total intensity
and polarization behaviour during propagation of the plasma modes in
the pulsar magnetosphere.  Nonetheless, high detection sensitivity
studies of single pulse polarimetric emission from normal pulsars
(rotation periods longer than $\sim$100 msec) have revealed several
details regarding the emission mechanism. In this article we have
attempted to highlight some of these observations that provide
credible evidence for the radio emission mechanism in normal pulsars
to be CCR from charge bunches. These charge bunches excite the
linearly polarized $t$ and $lt$ modes in the plasma that can detach
from the medium almost instantaneously after excitation, with very
little propagation effects in the pulsar plasma. The detected circular
polarization in the observed emission requires the presence an
additional plasma components in the form of positively charged iron
ions.  Such ions are naturally generated in the Partially Screened Gap
above the pulsar polar cap region. We find effective evidence for
Partially Screened Gap from observations of subpulse drifting. We also
discuss how the drifting behaviour constrain the surface magnetic
field to be non-dipolar in nature.

\section{Polarization Behaviour in Normal Pulsars} 

\subsection{The Rotating Vector Model}
A series of papers appeared in 1969 showing the linear polarization
position angle (PPA hereafter) to exhibit a characteristic S-shaped
sweep across the emission window of a pulsar. These results followed
from the radio observations of Vela Pulsar \cite{1969ApL.....3..225R}
as well as the optical emission from the Crab pulsar
\cite{1969ApJ...157L...1W,1969ApJ...156L.131B}. The rapid sweep of the
PPA strongly suggests a rotational model in pulsars and as a
consequence the `single-vector' or the `rotating vector' model (RVM)
was proposed where the position angle of the projected single-vector
changes due to rotation of the pulsar \cite{1969ApL.....3..225R}. The
rotation of the position angle has also been observed in Jupiter where
the electric field vectors due to synchrotron emission are
perpendicular to the magnetic field planes \cite{1962ApJ...136..276M}
suggesting that the RVM can be generally applied to all such systems
(see e.g. \cite{2019ApJ...887...44D}). The observations of the Vela
pulsar led \cite{1969ApL.....3..225R} to conclude that the radio
emission originates close to the dipolar magnetic field pole of the
neutron star. The strong magnetic field near the magnetic pole rules
out synchrotron emission as a possible mechanism leaving curvature
radiation from ultra-relativistic charges, moving along curved
magnetic field lines, as a viable mechanism
\cite{1969PASA....1..254R}. If the curvature radiation is triggered in
vacuum the electric field is expected to lie in the plane of the
magnetic field. There was also alternative ideas that the observed PPA
can be explained from the radiation source being located at the light
cylinder \cite{1969Natur.221...25G} or even significantly further away
from it \cite{1970ApJ...160.1003L}, see \cite{1975ARA&A..13..511G} for
a detailed review about these early conclusions). However, at these
distances the PPA features will be distorted by aberration-retardation
effects (see discussion below), thereby ruling out these
possibilities.

Generally, the PPA variations in pulsars have a S-shaped sweep
resembling the RVM, however, there exists a subset of pulsars with
extremely complex PPA behaviour (e.g. \cite{1975ApJ...196...83M}.
Detailed single pulse studies reveal that part of the complexity seen
in average PPA traverse arises due to the presence of orthogonal
polarization mode, OPM \cite{1975PASA....2..334M,
  1976Natur.263..202B}. In several cases the presence of
non-orthogonal PPA is also detected. The OPM makes the vacuum
curvature radiation model untenable in pulsars
\cite{1969PASA....1..254R}, with the ordinary and extraordinary mode
of the magneto-ionic plasma gaining favourability
\citep{1976ApJ...204L..13C, 1977PASA....3..120M}. The non-RVM like PPA
behaviour also gives rise to the possibility that radio emission in
some pulsars may originate from non-dipolar magnetic fields. But in
pulsars where RVM type of PPA traverse is found, clear
phenomenological association can be established between the shape of
the average profile with the PPA traverse. Using spherical geometry
the RVM of PPA traverse can be expressed in the form :
\begin{equation}
\Psi = \Psi_{\circ} + 
	\tan^{-1}{\left(\frac{\sin{\alpha} \sin{(\phi-\phi_{\circ})}}
{\sin{(\alpha + \beta)} \cos{\alpha} - \sin{\alpha} \cos{(\alpha + \beta)}\cos{(\phi-\phi_{\circ}}}\right)} \\
\label{eq1}
\end{equation}
here $\Psi$ is the PPA, $\phi$ is the rotational pulse phase, $\alpha$
is the angle between the rotation axis and the magnetic axis and
$\beta$ is the angle at closest approach between the rotation axis and
the observer line of sight (LOS). $\Psi_{\circ}$ and $\phi_{\circ}$
defines the reference points for measurement offsets.

The steepest gradient (SG) point of the PPA lies in the plane
containing the rotation and magnetic axis, and is related to $\alpha$
and $\beta$ as $d\Psi/d\phi = \sin(\alpha)/\sin(\beta)$. Eq.~\ref{eq1}
is a purely geometrical construction, and has no provision for
interpreting the OPMs (or non-OPMs). In several pulsars two parallel
PPA tracks can be found separated by 90$^{\circ}$ in PPA phase, and
the same RVM curve can be used for both OPM tracks. Hence, considering
RVM to be a purely geometrical model, it can be seen that large values
of $\mid d\Psi/d\phi \mid$ (i.e. low $\beta$) are associated with
profiles with multiple components, while small values of $\mid
d\Psi/d\phi\mid$ (high $\beta$) is usually found in single component
profiles. A statistical study comprising of a significant number of
pulsars can be used to construct a two dimensional structure of the
emission beam. The best approximation of the average emission beam is
believed to comprise of a nested core-cone structure, with a central
core emission around the magnetic dipole axis surrounded by two rings
of nested conal emission. The average shape depends on the LOS
traverse across the emission beam (see discussion in Section 3.1). A
number of studies have used the core-cone morphology to obtain a
classification scheme for pulsars
\citep{1983ApJ...274..333R,1990ApJ...352..247R,1993ApJ...405..285R,
  1993ApJS...85..145R}, and estimated the location of the radio
emission region to be below 10\% of the light cylinder radius,
consistent with expectations of RVM \cite{1969ApL.....3..225R}.

\subsection{Effect of Magnetospheric Plasma on PPA}
There are several effects that can change the nature of PPA in the
radio emission, like relativistic modifications introduced by
rotation, if the source is located close to the light cylinder
\cite{1973ApJ...183..977F, 1976ApJ...205..247F}, and bending of the
dipolar field lines due to sweep-back effects near the light cylinder
\cite{1985SvA....29...33S}. It has also been suggested that wave
propagation in the magnetospheric plasma can affect the PPA shape
\cite{1986ApJ...303..280B}. The shift in PPA phases due to aberration
retardation effects, arising due to co-rotation of the plasma, has been
considered by \cite{1991ApJ...370..643B} (BCW hereafter). Other
suggested modifications include the effect of return currents
\cite{2001ApJ...546..382H}, and corrections due to the visible point
effect, since the LOS is not along the tangent to the magnetic field
lines \cite{2004ApJ...609..335G, 2014PASA...31...39Y}. Amongst these,
the most prominent feature seen in the observed PPA behaviour is the
aberration retardation shifts described by the BCW model, where radio
emission is expected to arise from relativistically streaming plasma
along the open field lines, and the influence of co-rotating plasma on
the RVM is estimated. The particles move along the curved dipolar
field lines and due to co-rotation they follow a helical trajectory in
the inertial frame. This aberration retardation effect, also called
the delay-radius relation, introduces a first order shift in the PPA
by an amount $-3r_{em}/R_{lc}$, here $r_{em}$ is the radio emission
height, $R_{lc}=cP/2\pi$ is the light cylinder radius, $P$ is the
rotation period and $c$ the speed of light. The beaming aberration
advances the emission by further $r_{em}/R_{lc}$, and hence the total
shift between the PPA and the emission window is $\Delta \phi = 4
r_{em}/R_{lc}$. BCW suggested that $r_{em}$ can be estimated by
measuring $\Delta\phi$, defined as the difference between the average
pulse profile center and the SG point. This method has been
subsequently used to estimate the emission height in a large sample of
pulsars, and have constrained radio emission to originate below 10\%
of the light cylinder radius, i.e.  $\mathcal{R}=r_{em}/R_{lc} < 0.1$,
and matches the geometrical height estimates
\cite{1991ApJ...370..643B,1992ApJ...385..282P,1993ApJ...405..285R,
  1996A&A...309..481X,1997MNRAS.288..631K,1997A&A...324..981V,1998MNRAS.299..855K,
  2002ApJ...577..322M,2003A&A...397..969K,2004A&A...421..215M,
  2008MNRAS.391.1210W,
  2011ApJ...727...92M,2023ApJ...952..151M,2023MNRAS.520.4582P}.

At the radio emission heights the first order shifts proposed by BCW
dominate over the other distortions
\cite{1973ApJ...183..977F,1976ApJ...205..247F, 1985SvA....29...33S},
although the propagation effects in the magnetosphere
\cite{1986ApJ...303..280B} may introduce comparable changes in
$\Delta\phi$.  The two primary propagation effects are due to
refraction and mode coupling.  After the emission is generated at
$r_{em}$, as the waves travel through the distance of the polarization
limiting radius, $r_{pl}$, their polarization features can change
\cite{1979ApJ...229..348C,1986ApJ...303..280B}. Depending on the
mechanism the emission frequency, $\nu_{r}$, is initially either close
to or lower than the plasma frequency, $\nu_p$. As the wave propagates
outwards the plasma density reduces and the value of local plasma
frequency decreases.  At the height specified by $r_{pl}$, $\nu_r >
\nu_{\circ} = 2\sqrt{\gamma_p}~\nu_p$ (here $\nu_{\circ} =
\omega_{\circ}/2\pi$, $\nu_p = \omega_p/2\pi$ and $\gamma_p$ is the
average Lorentz factor of the plasma), and the emission detaches from
the magnetosphere as electromagnetic waves. The dispersion relation of
the $t$-mode has vacuum like characteristics in the medium and as a
consequence does not suffer refraction. The $lt_1$-mode is
sub-luminous with refractive index $n>1$, while the $lt_2$-mode
comprises of the super-luminous branch with $n < 1$, and hence both of
them can be subjected to refraction in the medium. The other
propagation effect of interest is mode coupling due to ``adiabatic
walking'' \cite{1979ApJ...229..348C}. The two natural modes of the
plasma propagate independent of each other, but during propagation one
of them can generate the other mode and change the intrinsic
polarization in a manner such that the polarization vector remains
either parallel or perpendicular to the $\vec{k}$-$\vec{B}$ plane.
Once the adiabatic walking condition is violated, the modes can couple
with each other and emerge as elliptically polarized waves. Thus both
refraction and adiabatic walking can cause a twist in the PPA, causing
deviation from the RVM. Using typical pulsar parameters BCW concluded
that the effect of refraction and mode coupling are small and does not
affect the delay-radius relation significantly. This is further
supported by the high levels of symmetry in PPA traverse with very
close to RVM like behaviour observed in a large number of pulsars. If
the radio emission detaches from the pulsar magnetosphere at
significantly larger heights, i.e. large $r_{pl}$, then due to the
co-rotating plasma, the PPA will not be symmetrical in nature. This
serves as compelling evidence for the coherent radio emission in
pulsars to be generated in the inner magnetosphere, where $r_{pl}$ is
small.

Additional details about the nature of polarization have emerged from
observations of the Vela pulsar. The X-ray observation of the pulsar
wind nebula has revealed the projection of the rotation axis on to the
sky plane \cite{2001ApJ...556..380H,2001ApJ...549.1111L}. The
projection of the rotation axis in the plane of the sky makes an angle
of $\sim 90^{\circ}$ with the projection of the PPA at the SG point,
implying that the electric field of the emerging waves is
perpendicular to the dipolar magnetic field line planes. If we ignore
the propagation effects, the same argument can be extended to the rest
of the PPA traverse plane, leading to the conclusion that the emerging
radiation is perpendicular to the dipolar magnetic field line planes.
The direction of proper motion of the Vela pulsar was found to be
along the rotation axis, suggesting that the absolute PPA plane is
orthogonal to the direction of motion of the pulsar. In other pulsars
the rotation axis cannot be identified, but it is possible to find the
direction of proper motion in many cases. The angle between the
direction of proper motion and the absolute PPA plane,
$\Delta\Phi_{PM-PA}$, has a bimodal distribution around 0$^{\circ}$
and 90$^{\circ}$ \cite{2005MNRAS.364.1397J,2007ApJ...664..443R,
  2012MNRAS.423.2736N,2013MNRAS.430.2281N,2015MNRAS.453.4485F}, and
can be interpreted as the polarization of the emerging radiation being
either parallel or perpendicular to the magnetic field line planes.
However, this conclusion is tentative and the distribution of
$\Delta\Phi_{PM-PA}$ can be easily influenced by several other effects
\cite{2005MNRAS.364.1397J,2013MNRAS.430.2281N}.

\subsection{Example of Polarization in Pulsar Radio emission}
\begin{figure}[H]
\includegraphics[width=8.5 cm]{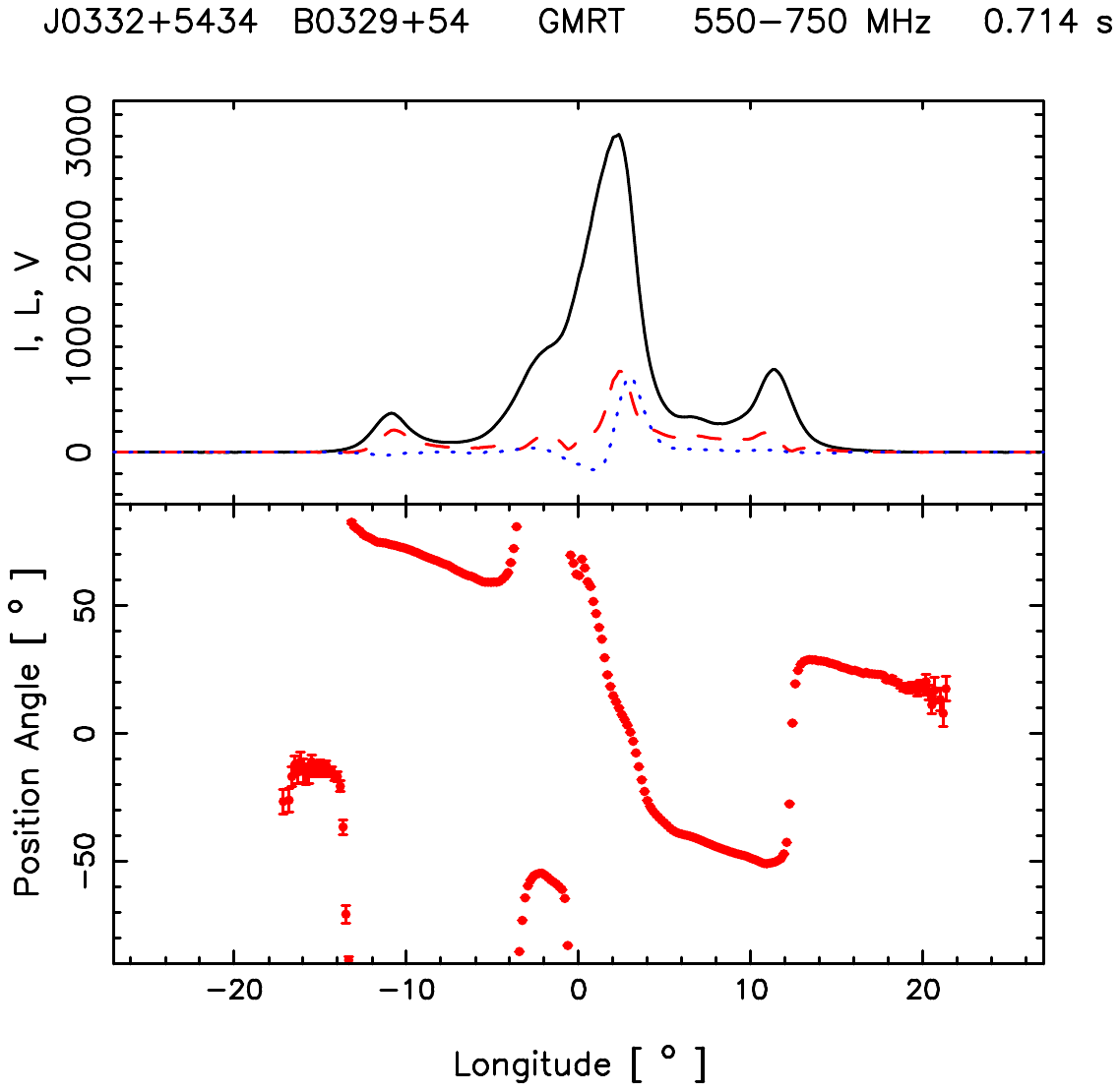}
\caption{The above plot shows the average polarization properties of PSR 
B0329+54 average over the wideband frequency range between 550 and 750 MHz. The
top panel shows the total intensity profile shape (black line), and the 
variation of the linear polarization (red dashed line) along with circular 
polarization (blue dotted line) across the profile. The bottom panel shows the
average PPA traverse that has quite complex features.
\label{fig1}}
\end{figure}   

The polarization behaviour in pulsars, as described in the previous
section, is illustrated by considering the example of PSR B0329+54
(J0332+5434). The parameters of this pulsar are $P = 0.715$ s, period
derivative $\dot{P}=2.05\times10^{-15}$ s~s$^{-1}$, characteristic age
$5.53 \times 10^6$ years, surface dipolar magnetic field
$1.22\times10^{12}$ G and spin-down energy loss of $2.22\times10^{32}$
ergs/s.  The average profile has five distinct components and can be
classified as Multiple (M) type (see Discussion in Section 3.1). PSR
B0329+54 is one of the brightest pulsars in the northern sky and has
been extensively studied over a wide frequency range. We observed the
pulsar using the Giant Metrewave Radio Telescope (GMRT), consisting of
30 antennas, with 14 antennas located within a central square
kilometer area, while the other 16 antennas are spread out in a
Y-shaped array with maximum distance of 27 km
\cite{1991CuSc...60...95S}. The observations were carried out in the
full polar, phased array mode of u-GMRT \cite{2017CSci..113..707G},
using the wideband receivers between frequency range of 550 and 750
MHz, and around 7000 single pulses were recorded with a time
resolution of 0.327 msec. The average profile is shown in
Fig.~\ref{fig1} along with the linear and circular polarization levels
across the profile window and the time averaged PPA traverse, that has
quite a complex nature unlike the RVM. In fact, in the normal pulsar
population the RVM does not fit the average PPA traverse in many
pulsars and detailed exercise of fitting the RVM to a large sample of
pulsars by \cite{2023MNRAS.520.4801J} showed that the RVM was a good
fit for 60\% of the pulsars and failed for about 40\%.

\begin{figure}[H]
\includegraphics[width=8.5 cm]{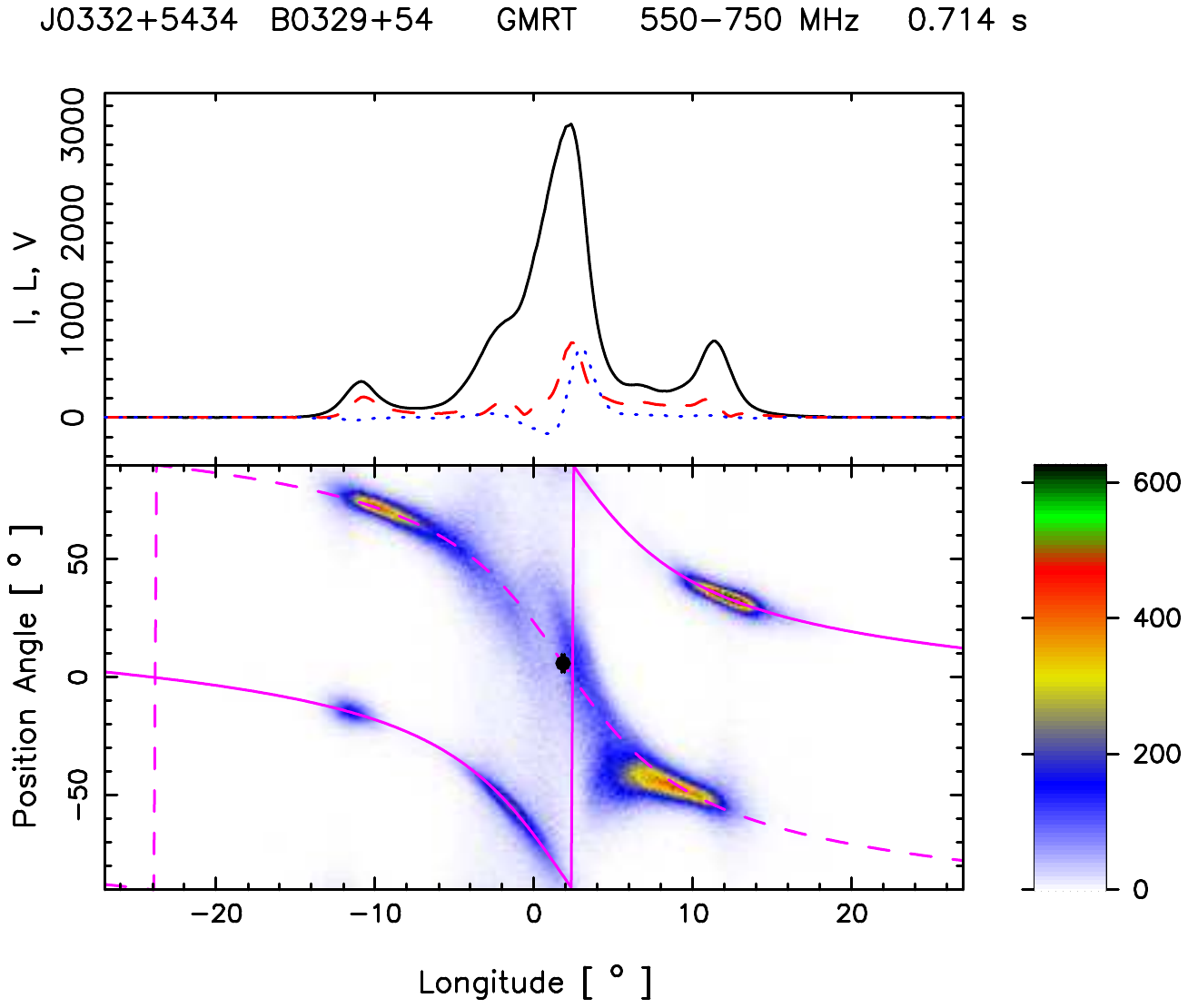}
\caption{The top panel shows the average profile of the total intensity (Stokes
I; solid black lines), total linear polarization (dashed red line), and 
circular polarization (Stokes V; dotted blue line) of PSR B0329+54 between 
550 and 750 MHz observing frequency. The lower panel shows the single pulse
PPA distribution (color scale) along with the RVM fits to the two PPA tracks, 
shown as dashed and solid magenta lines. 
\label{fig2}}
\end{figure}

The complex nature of the average PPA traverse of PSR B0329+54 was
first unraveled by \cite{1995MNRAS.276L..55G}, as they overlayed the
PPAs of single pulses to find two clear tracks separated by
90$^{\circ}$, that correspond to the two OPMs, as well as other PPAs
outside the two tracks. Later studies at multiple frequencies
confirmed the detection of the two OPM in this pulsar
\cite{2007MNRAS.379..932M}. The single pulses also show the mode
changing behaviour, where the radio emission switches between two
stable states with different profile shapes
\citep{1971MNRAS.153P..27L,1982ApJ...258..776B,
  2011ApJ...741...48C,2018Ap&SS.363..110B}, but the PPA behaviour is
similar in the two emission modes \cite{2019MNRAS.484.2725B}.
Fig~\ref{fig2} shows the single pulse PPA distribution from the u-GMRT
observations, with the two PPA tracks clearly visible
\cite{1995MNRAS.276L..55G}.

Subsequent single pulse studies have demonstrated that PPAs of time
samples with high levels of linear polarization closely follow the RVM
nature
\cite{2009ApJ...696L.141M,2023MNRAS.tmpL..22M,2023ApJ...952..151M}.
Fig~\ref{fig3} shows a subset of the PPA behaviour of Fig~\ref{fig2}
where the time samples with linear polarization level in excess 90\%
have been selected.  The two PPA tracks of the OPMs are clearly
evident in this plot. Each PPA track can be fit with the RVM using the
same values of $\alpha$, $\beta$ and $\phi_{\circ}$, and tracks are
exactly orthogonal. \cite{2007MNRAS.379..932M} estimated the absolute
direction of the PPA with respect to the fiducial plane, containing
the rotation axis and the magnetic dipole, and identified the two
tracks as the X-mode (dashed magenta line in Fig~\ref{fig2}, lower
window) and the O-mode (solid magenta line in Fig~\ref{fig2}, lower
window). The radio emission height can be determined form these
observations by estimating the aberration-retardation shifts in the
PPA, as discussed in the previous section.  We found the shift to be
$\Delta \phi\sim 2^{\circ}$, and the corresponding emission height is
about 300 km. In Fig~\ref{fig3} it can be seen that in a narrow region
around the longitude $2^{\circ}$, which is also close to the SG point,
the average linear polarization level goes below 60\%. This is due to
the large spread of single pulse PPAs in this longitude range, and we
have ignored this narrow region while carrying out the RVM fits.
Fig.~\ref{fig4} shows a second subset of the PPA behaviour of
Fig~\ref{fig2} where the time samples with linear polarization level
less than 30\% have been selected. The non-orthogonal PPAs are seen in
this figure to be present across the profile.

\subsubsection{High Levels of Linear Polarization}
\begin{figure}[H]
\includegraphics[width=7.0 cm]{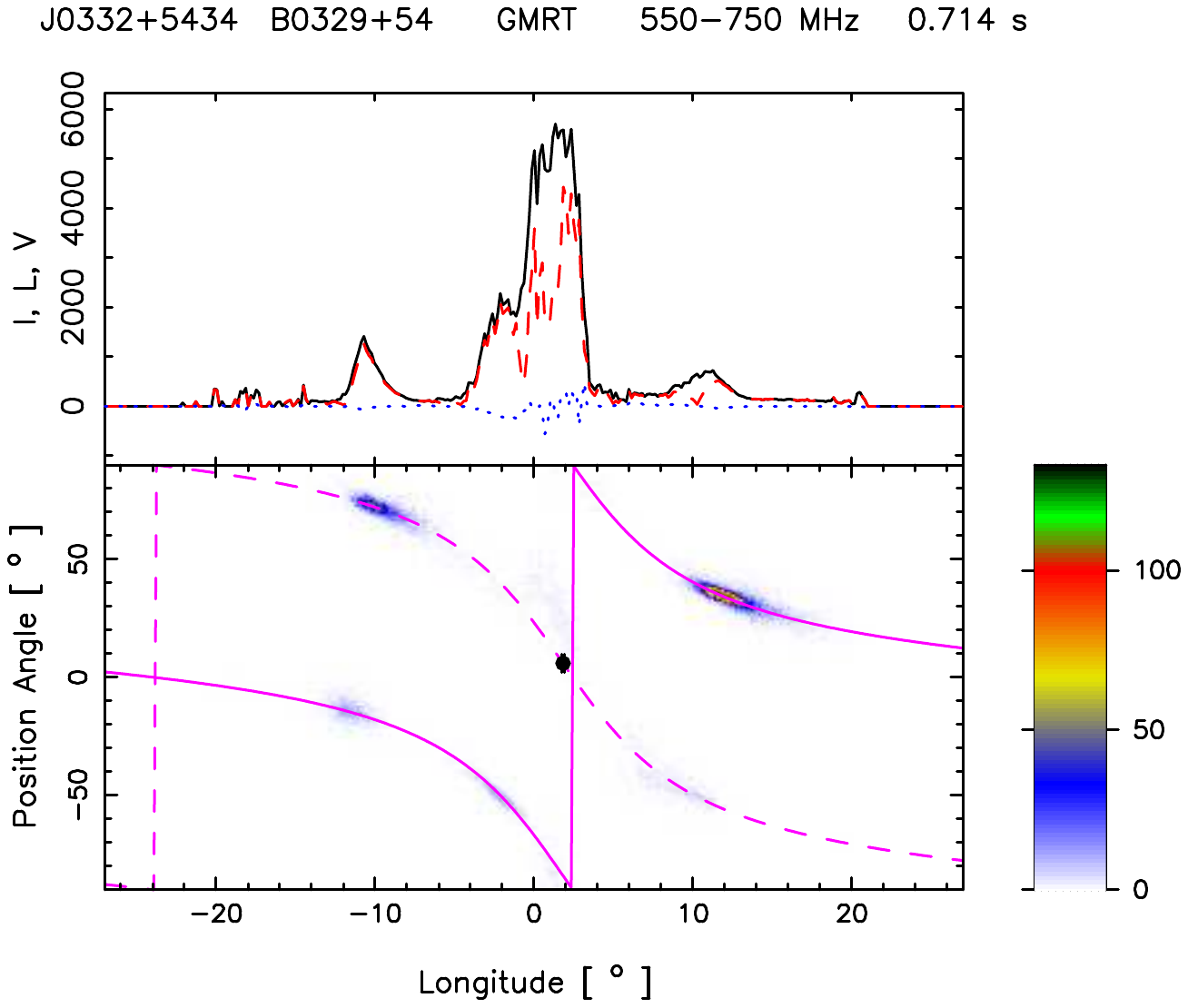}
\hspace{10px}
\includegraphics[width=6.5 cm]{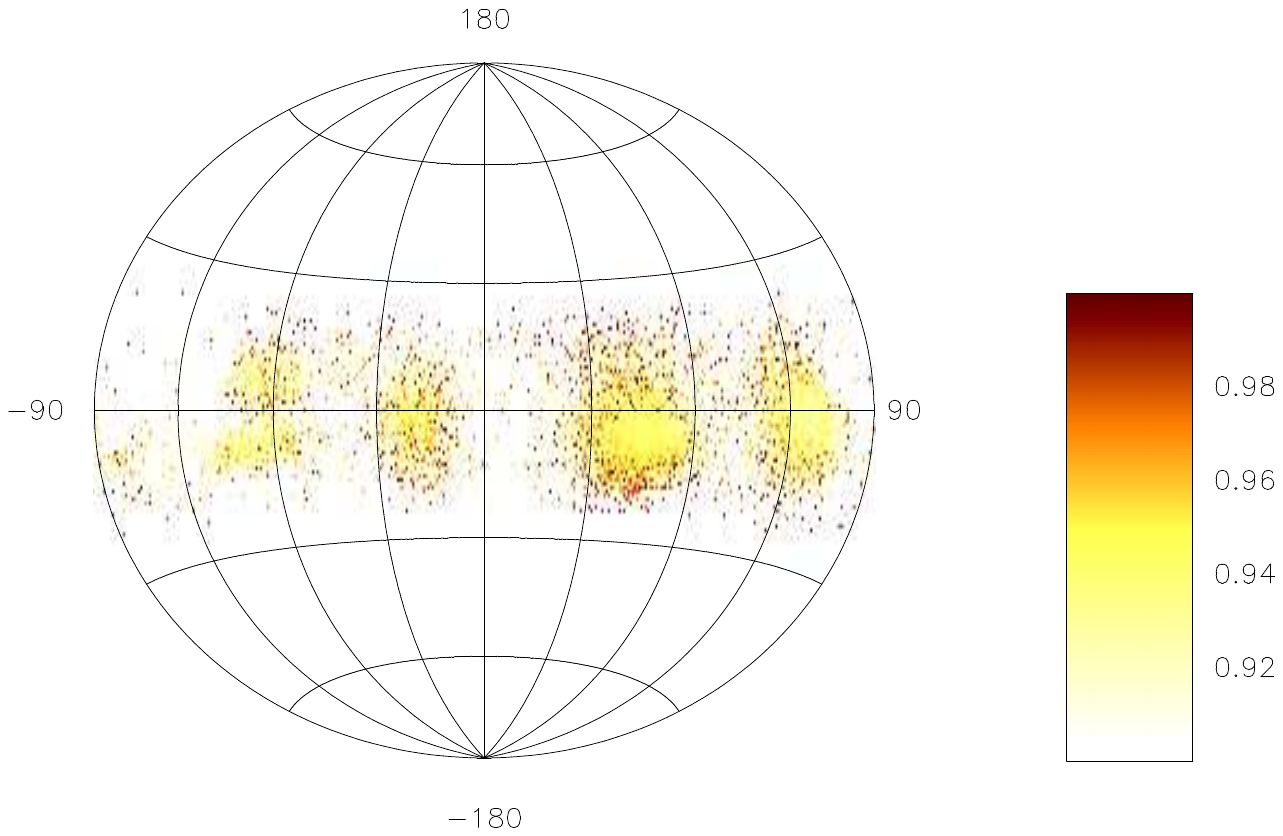}
\caption{The above plot shows a subset of the observed polarization behaviour 
in PSR B0329+54 for time samples with linear polarization level larger than 
90\%. In the left window, the top panel shows the average profile with total 
intensity (Stokes I; solid black lines), total linear polarization (dashed red 
line), and circular polarization (Stokes V; dotted blue line). The lower panel
in the left window shows the single pulse PPA distribution (color scale) with 
distinct orthogonal tracks. The RVM fits to the PPAs are also shown as dashed 
and solid magenta lines in this plot. The right plot show the Hammer$-$Aitoff 
projection of the polarized time samples with the color scheme representing the
fractional polarization level.
\label{fig3}}
\end{figure}

\begin{figure}[H]
\includegraphics[width=7.0 cm]{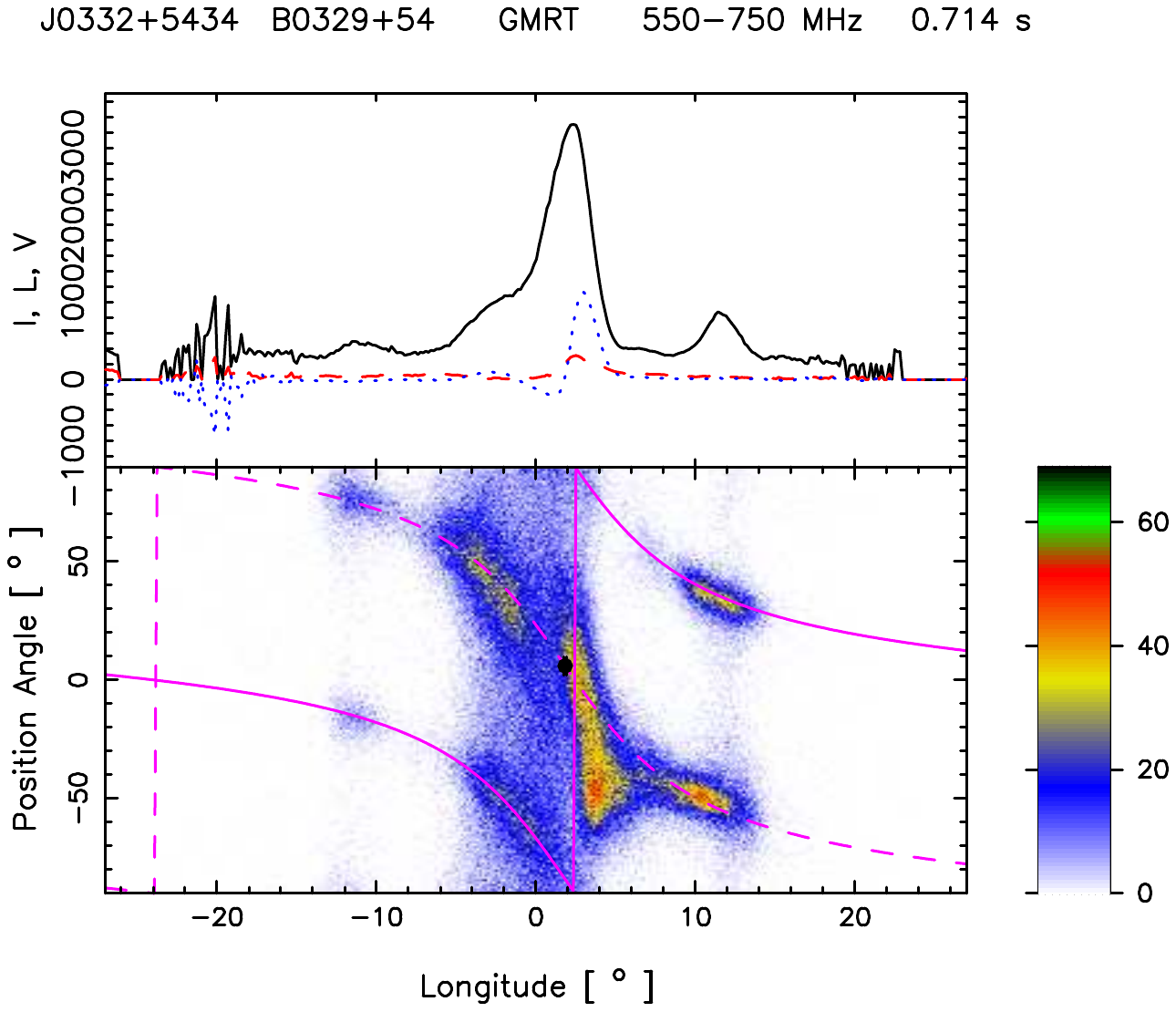}
\hspace{10px}
\includegraphics[width=6.5 cm]{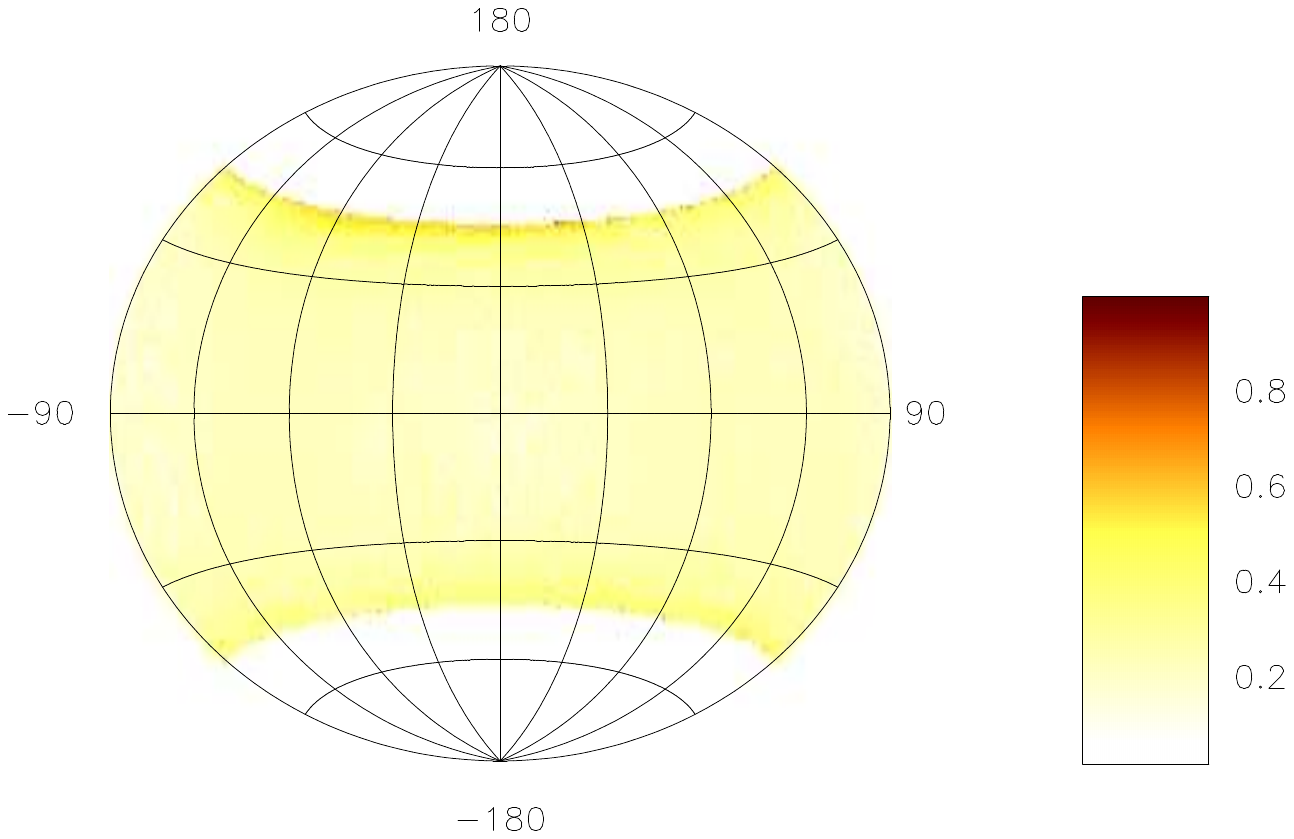}
\caption{The above plot shows a subset of the observed polarization behaviour 
in PSR B0329+54 for time samples with linear polarization level less than 30\%.
In the left window, the top panel shows the average profile with total 
intensity (Stokes I; solid black lines), total linear polarization (dashed red 
line), and circular polarization (Stokes V; dotted blue line). The lower panel 
in the left window shows the single pulse PPA distribution (color scale) with 
non-orthogonal characteristics. The RVM fits to the PPAs are also shown as 
dashed and solid magenta lines in this plot. The right window shows the 
Hammer$-$Aitoff projection of the polarized time samples with the color scheme 
representing the fractional polarization level. 
\label{fig4}}
\end{figure}

The presence of individual subpulses with high levels of linear
polarization was detected in several earlier studies
\cite{1969Natur.223..934S, 1976QJRAS..17..383S}, but their origin was
unclear. If we consider the situation in Fig~\ref{fig3}, where time
samples with close to 100\% linear polarization levels are shown, the
two PPA tracks have identical RVM characteristics with a $90^{\circ}$
shift in PPA phase. Although the RVM was proposed for vacuum curvature
radiation originating close to the poles of a static magnetic dipole,
the two PPA tracks represent plasma effects, and can be associated
with the natural plasma modes, i.e. $t$-mode and $lt$-mode of the pair
plasma (see Discussion is section 2.2). However, the $t$-mode and
$lt$-mode are confined to the plasma and additional details are
required to understand how they transform into the electromagnetic X
and O-modes that can escape the plasma, and yet retain the vacuum like
PPA characteristics.

The $t$-mode and $lt$-mode have slightly different dispersion
relations in the magnetospheric plasma and can propagate independent
of each other. It has been noted by \cite{1979ApJ...229..348C} that in
a slowly varying magnetoionic inhomogeneous plasma the polarization
features at the emission region evolves due to adiabatic walking. At a
certain distance the effect of adiabatic walking is terminated and the
observed polarization behaviour reflects the pattern frozen in at this
point. Under these circumstances it is possible to obtain high levels
of linear polarization, but the effect of emission mechanism on the
PPA is lost. On the contrary the single pulse observations reported in
\cite{2009ApJ...696L.141M} showed that subpulses with high levels of
linear polarization had PPAs across them that closely followed the RVM
nature. Also in a recent study \cite{2024MNRAS.tmp.1181J} shows that
by applying the high polarized time sample criteria, the RVM can be
recovered for a large number of pulsars and for several pulsars the
high polarized samples follow two parallel RVM like PPA tracks. These
observations hence strongly favour the natural modes in the plasma to
be excited by CCR, with the radiation detaching from the magnetosphere
almost immediately after they are generated, with little influence of
propagation effects like adiabatic walking \cite{2014ApJ...794..105M}.
Given the fundamental importance of propagation effects on the
observed polarization behaviour and their ability to uncover the
emission mechanism, we discuss below the possibility in more detail.

\section*{Propagation Effect Cannot Explain the High Levels of Linear Polarization}
The aberration and retardation effect is likely to affect both the
$t$- and $lt$-modes propagating in the plasma, and the effect depends
on the emission height from which each mode detaches from the plasma.
The $t$-mode, with vacuum like dispersion property, remains unaffected
by refraction in the medium and propagates along the tangent to the
magnetic field line at the point of generation. On the other hand both
branches of the $lt$-mode are refracted in the medium and are ducted
along the magnetic field lines with different extents. If the $t$-mode
and the $lt$-mode detach from the magnetosphere at different heights
the observed PPA tracks should be different.  However, the PPA tracks
of the two OPM are observed to be identical, suggesting that the
effects of refraction is minimal. As a result we can conclude that the
two modes escape the plasma from the same height.

\begin{figure}[H]
\includegraphics[width=10.5 cm]{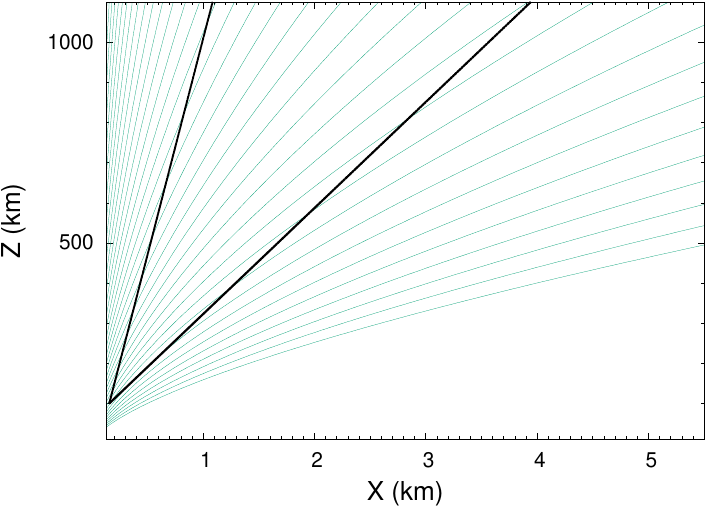}
\caption{The above plot shows the schematic of the open dipolar magnetic field 
lines (in green) above the neutron star surface. The black diverging lines 
correspond to the rectilinear propagation vector of an $1/\gamma_p$ emission cone
of $t$-mode generated at a height of 100 km from the surface, where $\gamma_p = 350$. The radio emission 
detaches around 1000 km from the surface. 
\label{fig5}}
\end{figure}

In order to understand the effect of adiabatic walking on the radio
emission let us assume the presence of a hitherto unknown mechanism
that is capable of exciting the normal modes in the relativistic pair
plasma. The emission is confined within a narrow cone ($\sim
1/\gamma_p$) centered around the tangent to local magnetic field. The
excitation frequency is close to the plasma frequency and the linear
polarization is either parallel or perpendicular to the
$\vec{k}\times\vec{B}$ plane\footnote{For example the linear
  acceleration mechanism excites the O-mode and has polarization
  oriented parallel to the $k-B$ plane}, but when averaged within the
cone the emission is completely depolarized. The longitude resolution
in Fig.~\ref{fig3} is 0.16$^{\circ}$, and if associated with the
angular size of the relativistically emitting cone, the corresponding
Lorentz factor is $\gamma_p\sim350$. A mechanism to obtain close to
100\% linear polarization from the $1/\gamma_p$ emission cone, based
on the adiabatic walking condition, was proposed by
\cite{1979ApJ...229..348C}. As the radiation propagates, the curved
magnetic field lines bend and due to adiabatic walking the emission
cone is no longer centered around the tangent to the local field
lines, but is directed away from them in its entirety. The
polarization vector needs to adjust itself to remain parallel or
perpendicular to the $\vec{k}\times\vec{B}$ plane, and hence in this
situation all the polarization vectors within the cone are pointed in
the same direction. Hence, the resultant polarization within the
emission cone can reach levels close to 100\%. This behaviour is
illustrated in Fig. 1 of \cite{1979ApJ...229..348C}, and shows the
reorientation of the polarization vector in the emission cone due to
adiabatic walking. However, one possible issue in this representation
is that the cross-section of the emission cone is assumed to be same
at different heights. On the contrary, the cross-section of the cone
keeps on increasing as the emission propagates upwards, as shown in
Fig.~\ref{fig5} for the $t$-mode generated at a height of 100 km above
the surface. The black lines in the figure correspond to a
$1/\gamma_p$ cone, with $\gamma_p \sim 350$, while green line
represent the open magnetic field lines. The radio emission detaches
from the plasma at a height of $\sim$1000 km, where the diverging
field lines are still enclosed by the $1/\gamma_p$ cone. Any averaging
effect within this cone would likely result in complete depolarization
of the emission \cite{2014ApJ...794..105M}.  Similar arguments can be
made for the $lt$-mode as well.

Thus we conclude that the time samples with high levels of linear
polarization seen in Fig.~\ref{fig3} cannot be explained by adiabatic
walking. Additionally, the symmetrical nature of the PPA traverse
requires extreme fine tuning of the plasma parameters. The multiple
components in the profile also suggests the presence of non-stationary
plasma flow, and cannot satisfy the conditions for adiabatic walking,
i.e. slow change of plasma density.  {\bf It should be noted that in
  average profile the fractional polarization across a wide frequency
  range can be frequency dependent, where higher frequency tends to
  have more depolarization (see e.g.
  \cite{1970ApL.....5..181M,1973ApJ...179L...7M,1996A&A...309..481X}).
  Generally depolarization is attributed to the mixing of OPMs, and
  some studies has been devoted to the mechanisms of depolarization
  (\cite{1998ApJ...502..883M,2005MNRAS.359..481K}), and also their
  frequency dependent behaviour (\cite{2015MNRAS.448..771W}). In this
  context simultaneous wide band observations of high linearly
  polarized time sample in order to explore their frequency dependence
  can be quite instructive.}

\section*{High Levels of Linear Polarization from Curvature Radiation}
Curvature radiation excites the $t$ and $lt$ modes, and due to
difference in refractive indices the modes split in the plasma medium
and can travel independently as linearly polarized modes,
perpendicular and parallel to the curved magnetic field line planes,
respectively \cite{2004ApJ...600..872G}. There is evidence for the
observed PPAs from normal pulsars being either parallel or
perpendicular to the curved magnetic field plane. If CCR indeed acts
as the radio emission mechanism in pulsars, then several critical
conditions related to coherence and escape of waves need to be
satisfied. \cite{1975ApJ...196...51R} suggested that linear
electrostatic Langmuir waves could be a possible source of CCR, but
this model was not found to be applicable by
\cite{1986FizPl..12.1233L} (see details in Paper II). An alternative
theory has been developed where it was suggested that the modulational
instability of Langmuir wave can form charge envelopes, or solitons,
that are stable enough to emit CCR
\cite{2000ApJ...544.1081M,2022MNRAS.516.3715R}.

\begin{figure}[H]
\includegraphics[width=13.5 cm]{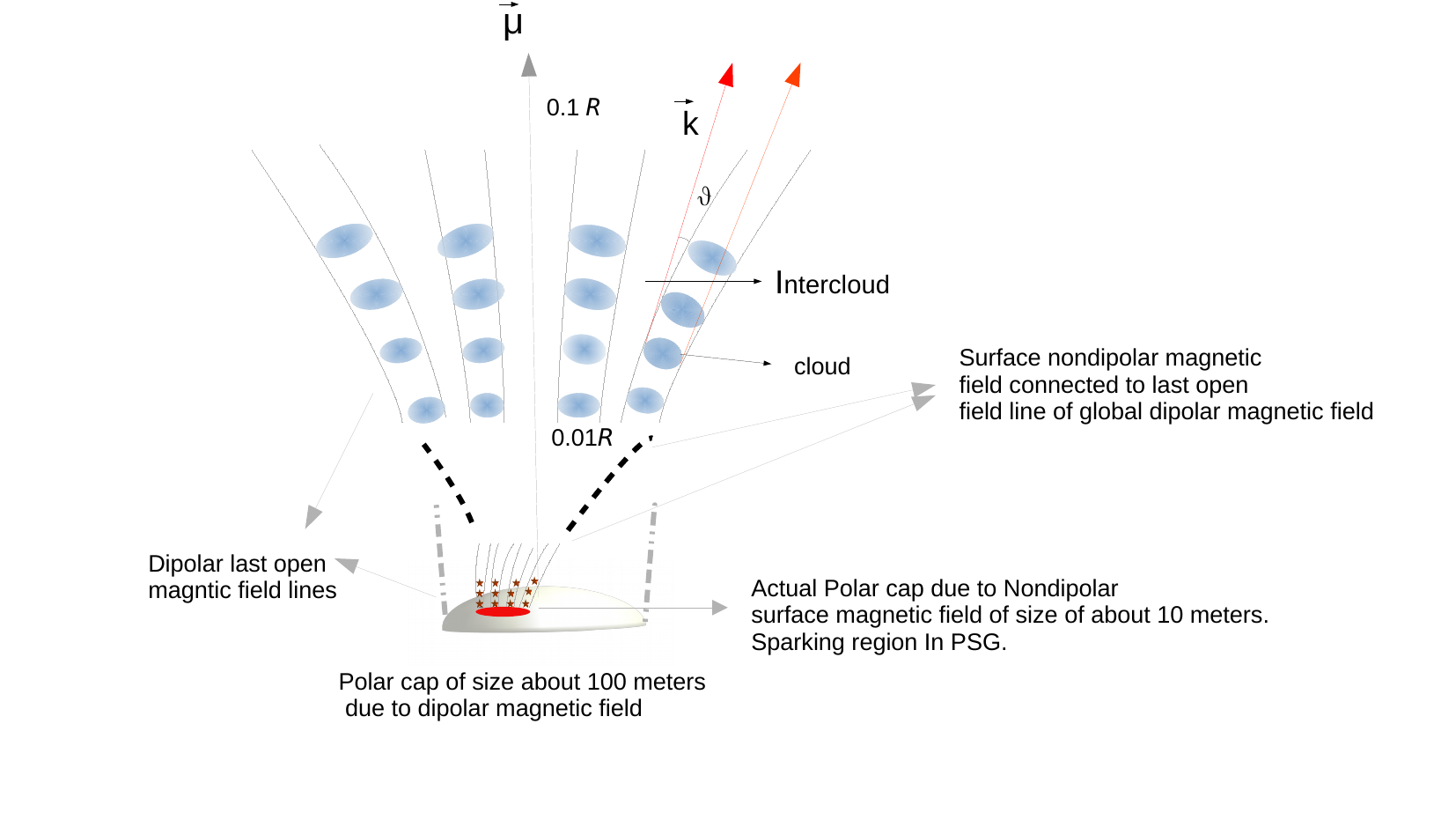}
\caption{The above plot shows a schematic of the PSG model (not to scale). Due 
to non-dipolar magnetic fields on the surface the actual polar cap (shown as 
red region) is smaller than the equivalent dipolar case. The PSG exists above 
the polar cap where non-stationary spark discharges are triggered, shown as 
red stars in the figure. The open field line region corresponding to the 
non-dipolar field lines connects to the global dipolar field (shown as dashed 
black line), and the spark discharges forms plasma clouds with high density regions near the center 
(blue shaded) and low density region inbetween. CCR is generated in the plasma clouds at about 
$\mathcal{R}\sim 0.01$ and the radiation enters the inter-cloud region below 
$\mathcal{R}\sim 0.1$. The inter-cloud regions are the white spaces enclosed 
within the open magnetic field line region where the density of the particles
injected from the adjoining sparks become lower and they are dominated by 
$^{56}Fe_{26}$ ions.  
\label{fig6}}
\end{figure}

\cite{1975ApJ...196...51R} also proposed the existence of an inner
vacuum gap region above the polar cap where non-stationary plasma flow
arises due to intermittent spark discharges and pair plasma cascade,
which eventually gives rise to a two stream instability. This model
can be used to address a number of observational phenomena such as
subpulse drifting, microstructures in the single pulses, and
measurements of hot polar cap. However, the vacuum gap model do not
quite agree with the details of these measurements and requires
modification in the form of a partially screened gap, PSG hereafter
\citep{2003A&A...407..315G} (see discussion below in
section~\ref{sec4}). An illustration of the non-stationary plasma flow
from the PSG model is shown in Fig.~\ref{fig6}. In the dense pair
plasma clouds stable relativistic charged solitons can be generated at
distances of few percent of $R_{lc}$, that have Lorentz factor
$\Gamma$, larger than the Lorentz factor of spark-associated plasma
$\gamma^{sp}_{s}$.  The solitons while moving along the curved
magnetic field lines excite CCR with characteristic frequency, $\nu_r
\sim 1.2 c \Gamma^3/2\pi \rho \sim 0.8\times
10^{-9}(\Gamma^3/P)\mathcal{R}^{-0.5}$, where $\rho \sim 2 (r_{em}
R_{lc})^{0.5}$ is the radius of curvature of the open dipolar field
lines.  If we use the parameters of PSR B0329+54, and assume typical
values of $\mathcal{R} \sim 0.01$, $\Gamma \sim 300$ we get $\nu_r
\sim 0.4$ GHz, which lies in the typical radio frequency range. Using
multiplicity factor of the spark-associated plasma, $\kappa^{sp} \sim
10^4$, the characteristic frequency $\nu_{\circ} \sim
2.4\times10^{11}~s^{-1}$ satisfies the condition $\nu_r \ll
\nu_{\circ}$ in PSR B0329+54.

The curvature radiation excites the sub-luminal $t$ and $lt_1$ modes
with polarization directed perpendicular and parallel to dipolar
magnetic field line planes, respectively. In the case of curvature
radiation emitted by individual particles under vacuum like
conditions, the power of the perpendicular component is seven times
lower than the parallel component. The two modes propagate
independently in the plasma medium, with the $t$-mode having vacuum
like properties and $lt_1$-mode getting ducted along the magnetic
field lines.  It has been shown that despite the power of the $t$-mode
of curvature radiation in plasma being suppressed, the emitted
intensity is sufficient to account for the observed pulsar
luminosity \cite{2004ApJ...600..872G}. In order to estimate the
rotation of the polarization plane of the modes, the adiabatic walking
condition needs to be evaluated, and has the form
\cite{1979ApJ...229..348C}
\begin{equation}
\mid \frac{1}{k}\frac{\partial}{\partial x} \psi \mid \ll \mid \Delta N \mid
\label{eq1}
\end{equation}
Here $N$ is difference between the refractive index of $t$ and $lt_1$
mode, $\Delta N$ is the change in $N$ as the wave propagates in the
medium, $k$ is the wave vector and $\psi$ is the linear PPA. In case
of curvature radiation from solitons in the radio frequency range, the
condition $\nu_r \ll \nu_{\circ}$ is satisfied and $\Delta N \sim 0$,
such that the effects of adiabatic walking is not
applicable\footnote{Note that \cite{1979ApJ...229..348C} found
  adiabatic walking to be important as they were considering an
  emission mechanism where the radio frequency is close to the
  characteristic frequency, i.e. $\nu_r \approx \nu_{\circ}$, and
  $\Delta N$ has large values.}
\cite{2004ApJ...600..872G,2014ApJ...794..105M}.  The waves preserve
their polarization direction while escaping from the plasma (also see
\cite{2022MNRAS.512.3589R}). The $lt_1$ mode cannot escape under
homogeneous plasma conditions as it gets ducted and eventually
undergoes Landau damping. However, $\nu_r > \nu_{\circ}$ is possible
if the plasma density has large gradients, and then the wave can
successfully escape from the plasma. In the PSG model the center of
the dense clouds have high pair multiplicity factor, $\kappa^{sp}$,
while as one moves towards the boundary between adjacent clouds the
muliplicity factor, $\kappa^{isp}$, decreases to orders of unity. This
creates a density gradient in the plasma at the inter-cloud boundary
through which the $lt_1$ mode can escape \cite{2014ApJ...794..105M,
  2023ApJ...952..151M}.

A large number of solitons are produced in each plasma cloud and as
the cloud moves along the curved field line the average emission from
the cloud passes into the inter-cloud region. The radiation propagates
in a rectilinear manner in the inter-cloud region, and due to the
curvature of the field lines the angle between $\vec{k}$ and the
tangent to the dipolar magnetic field, $\vartheta$, continues to
increase. The plasma clouds are expected to give rise to the
individual components seen in the average profile, with typical
angular width of about $2^{\circ}$. CCR is generated at heights of
$\mathcal{R} \sim 0.01$ where $\vartheta \sim 0^{\circ}$ and as the
radiation reaches $\mathcal{R} \sim 0.1$ the angle $\vartheta$
increases to $\sim4^{\circ}$ which is large enough for the radiation
to transfer into the inter-cloud region \cite{2023ApJ...952..151M}.
Beyond this region $\vartheta$ does not change with increasing height
till $\mathcal{R} \sim 0.5$, after which the magnetic field lines bend
due to sweepback effects. The $t$ and $lt_1$ modes from different
plasma clouds enter the inter-cloud region and undergo incoherent
addition. This process gives rise to the resultant linear polarization
and the associated PPA distribution observed in pulsars. The emission
process is stochastic with different clouds having different
orientations of the $\vec{k} \times \vec{B}$ plane, such that
averaging will lead to depolarization. From time to time we expect
dominant $t$ or $lt_1$ modes from strongly magnetized plasma to emerge
from the clouds, and the corresponding X- and O-modes will show up as
the clear orthogonal PPA tracks observed in Fig.~\ref{fig3}. At other
times the emergent emission undergo incoherent mixing of the $t$- and
$lt_1$-modes in various proportions and orientation, resulting in the
low levels of linear polarization and scattered PPAs seen in
Fig.~\ref{fig4}. The plasma clouds can also be associated with the
quasi-periodic structures, known as microstructures, observed in
single pulses \cite{1990AJ....100.1882C,2015ApJ...806..236M}.

CCR excited in the plasma clouds and subsequently emerging from the
inter-cloud region provides a successful mechanism to explain the
highly polarized time samples observed in pulsars, but this model
still remains largely qualitative. It has also been suggested that in
homogeneous plasma the cyclotron damping will affect the plasma modes
in the upper magnetosphere, preventing the radiation from escaping
\cite{1982SvAL....8..369M}. However, the inhomogeneities introduced by
the PSG model will reduce the damping effect as the waves propagate in
the inter-cloud region. Detailed theoretical studies are still
necessary to find the escape condition of $lt_1$ mode, as well as
simulations showing that the incoherent averaging process is able to
reproduce the observed polarization characteristics. Our discussions
make it clear that propagation effects due to refraction and adiabatic
walking have negligible effect in modifying the radiation features.

\subsubsection{Circular polarization}
In early studies of pulsar emission the circular polarization was
expected to be a part of the emission process
\cite{1984ApJ...277..367B,
  1990A&A...234..237G,1990A&A...234..269G,1990ApJ...352..258R,
  1987ApJ...322..822M,1997A&A...327..155G}, although the polarization
behaviour is likely to change as the emission propagates within the
plasma medium. The circular polarization behaviour of highly polarized
time samples from PSR B0329+54 is shown in Fig.~\ref{fig3}, where the
right window represents the Hammer-Aitoff projection of the
Poincar\'{e} sphere. The spread of the distribution away from the
equatorial plane indicates that almost all polarized time samples are
elliptically polarized. This suggests that the linearly polarized $t$-
and $lt$-modes generated in the inner magnetosphere are partially
converted into circularly polarized modes as they propagate further in
the outer magnetosphere. The polarization behaviour of PSR B0329+54
corresponding to the time samples with lower levels of linear
polarization is shown in Fig.~\ref{fig4}. The distribution spans a
wider azimuthal range on the Poincar\'{e} sphere with similar values
for both the positive and negative circular polarization. In the
entire observing span, shown in Fig.~\ref{fig2}, the mean level of
absolute circular polarization is about 20\%, which is typical for
normal pulsars \cite{2023ApJ...952..151M}. The models of emission
mechanism in pulsars need to explain the observed circular
polarization that arises due to propagation effects within the plasma
at heights between $\mathcal{R} \sim 0.01$ and $\sim0.1$.

One of the suggested mechanism for the appearance of circular
polarization is the wave mode coupling that appear due to co-rotation
of the magnetosphere
\cite{1979ApJ...229..348C,1986ApJ...303..280B,2000A&A...355.1168P,
  2010MNRAS.403..569W}. The propagation vector gets tilted with
respect to magnetic field line due to co-rotation of the magnetosphere
and as it travels a distance $r_p$, the adiabatic walking condition
adjusts the polarization to generate two independent modes, parallel
and perpendicular to the $\vec{k} \times \vec{B}$ plane. Eventually
the adiabatic walking is no longer dominant, resulting in the two
modes getting coupled to emerge as elliptically polarized waves. The
resultant sense of circular polarization is expected to be constant
across the profile, but in reality both sense, i.e. right and left
circular polarization are observed across the profile. To address this
discrepancy it was suggested that change in the plasma distribution
along the open flux tube can bring about the asymmetric nature in the
profiles \cite{2000A&A...355.1168P}. But as asserted in the previous
section, we have ruled out the possibility of adiabatic walking
condition to explain the observed high the high linear polarization
signals.  However, the effects of adiabatic walking and wave coupling
on the circular polarization needs further study using realistic model
of non-stationary plasma flow.

We consider another possibility where the generation of circular
polarization requires breakdown of symmetry in the positive and
negative components of the pair plasma. It should be noted that the
circular polarization naturally occurs as eigenmodes in a electron-ion
plasma. The co-rotational electric field causes the distribution
function of electrons and positrons to be slightly different and the
presence of ionic species can further break the gyrotropy. In this
case the normal modes propagating at small angles with respect to the
external magnetic field can be circularly polarized
\cite{1982PASA....4..365A,1991MNRAS.253..377K, 2010MNRAS.403..569W}.
The absence of gyrotropy has been used to explain the origin of
circular polarization using the PSG model \cite{2023ApJ...952..151M}.
It is convenient to consider a co-ordinate system
\citep{1967RvPP....3....1S} with the z-axis along, and introducing
$a_{x} = E_x/iE_y$, $a_{z} = E_z/iE_y$, i.e.
$\vec{E}=E_{y}(ia_{x},1,ia_{z})$, we can obtain the dispersion
relation as:
\begin{equation}
a_{x}^{2}-\Theta a_{x}-1 = 0. 
\label{3}
\end{equation}
The solution of equation (\ref{3}) has the form
\begin{equation}
a_{x}=\frac{1}{2}\Theta \pm \sqrt{\left( \frac{1}{2}\Theta \right) ^{2}+1}.
\end{equation}
Here $\Theta$ depends on the plasma parameters. When $\Theta \ll 1$
then $a_{x}\simeq \pm 1$ resulting in circular polarization. In the
PSG model the emission enters the ion dominated inter-cloud region at
a distance $\mathcal{R}\gtrsim0.01$. In the inter-cloud region, where
$\kappa^{isp} \sim 1$ and $\mathcal{R}\sim0.1$, the radio frequency is
higher than the local characteristic frequencies.  The wave can
propagate rectilinearly and the angle between the wave vector and the
local field, $\vartheta$, remains constant for large distances,
satisfying the conditions $\sin\vartheta \gg 1/\gamma_s^{isp}$ and
$\sin{\vartheta} < 1/\gamma_{ion}^{isp}$. It has been shown
\citep{2023ApJ...952..151M} $\mid \Theta \mid$ depends on the pulsar
parameters as:
\begin{equation}
\mid \Theta \mid = \cfrac{0.087}{\sin^2(\vartheta)}~\mathcal{R}^3 \cfrac{\kappa^{isp}}{{(\gamma _{s}^{isp})^3}} \left(\cfrac{P^5}{\dot{P}_{-15}}\right)^{0.5} \omega.
\end{equation}
Here $\gamma_{s}^{isp}$ is the average Lorentz factor of the electron
positron component of the plasma, and $\omega = 2\pi\nu$ where $\nu$
is the radio frequency in Hz. Using parameters for PSR B0329+54, and
other typical plasma parameters, $\vartheta \sim 0.07$ radian,
$\gamma_{s}^{isp}\approx300$, $\kappa^{isp}=1$ and $\nu=300$ MHz, we
estimate $\mid \Theta \mid \sim 0.37$.  This suggests that the above
mechanism supports the necessary condition for obtaining circular
polarization, i.e. $\mid \Theta \mid \ll 1$, in PSR B0329+54. The
observed circular polarization is an incoherent addition from many
different clouds, while the sense of circular polarization depends on
the viewing geometry.

\section{Variation of Spectral Index across the Profile Window}
The previous sections have highlighted that the magnetic field has
dipolar characteristics at the radio emission heights. The curvature
radiation is directly associated with the curvature of field lines. As
the LOS traverses across the emission beam, corresponding to the open
dipolar field line region, at any given height the radius of curvature
increases from the edge towards the center of the emission beam. The
shape of the average profile is determined by the LOS traverse across
the emission beam, with distinct physical properties seen in the core
component at the center compared to the conal components at the
periphery. The CCR from charge bunches predicts a difference in the
spectral nature of the flux density between the central core and the
peripheral conal components. A proper characterisation of this
spectral behaviour provides direct evidence for the emission mechanism
in pulsars. Below we summarize the nature of the average emission beam
in the pulsar population and the different efforts in constraining the
spectral variation across the emission beam.

\subsection{Classification of Profile Morphology and Nature of Emission Beam}
The average profile of each pulsar has unique features and
characterised by the number of components that can vary between one
and five. The components are classified into two major types, the core
component that is usually located at the center of the profile and
show large change of the PPA traverse across it with sign changing
circular polarization, and the conal components located around the
core with relatively shallow PPA slopes and higher levels of linear
polarization \citep{1983ApJ...274..333R} (also see discussions above).
The pulsar profile also shows significant evolution with frequency.
The profile widths become wider at lower frequencies due to the effect
of radius to frequency mapping, where the lower frequencies originate
from higher up the open dipolar field \citep{2002ApJ...577..322M}. In
addition the number of components in the profile can also change at
different frequencies.

The earliest suggestion about the nature of the emission beam was the
``hollow cone'' model \citep{1969Natur.221..443R,1976ApJ...209..895B}
where the emission was expected to originate in a narrow cone
surrounding the last open field lines, symmetrically around the
magnetic axis. However, the hollow cone model was not adequate to
explain all observed profile types and has led to the ``Core-Cone''
model of the emission beam proposed in a series of works
\citep{1983ApJ...274..333R,1990ApJ...352..247R,1993ApJ...405..285R,
  1993ApJS...85..145R,1993A&A...272..268G}, that has the widest
applicability in understanding the different profile types. In the
core-cone model the emission beam at any observing frequency comprises
of a central core region surrounded by two pairs of nested conal rings
forming the inner and the outer cones The profile shape is determined
by the pulsar geometry and the distance of the LOS traverse from the
magnetic axis. When the LOS cuts the emission beam centrally, close to
the magnetic axis, they form the core-cone Triple (T), and core-double
cone Multiple (M) profile types. Sometimes one side of the conal
component in a T profile is weaker and the profile is classified as
T$_{1/2}$. The conal Quadruple ($_c$Q), conal Triple ($_c$T), Double
(D) and conal Single (S$_d$) profiles represent the cases of the LOS
cuts being progressively away from the magnetic axis missing the
central core component and moving towards the edge of the beam. A core
Single (S$_t$) represents a central LOS traverse where the conal
components are absent. In many cases the S$_t$ profiles develop conal
outriders at higher frequencies, while the S$_d$ profiles becomes
wider and bifurcates at lower frequencies.

In the initial works the core-cone model preserved the hollow-cone
structure, with the outer cones and inner cones both bordering the
outer field line originating at different heights, while the core was
expected to arise from the surface. Subsequent works
\citep{2011MNRAS.414.1314M,2011MNRAS.417.1444M,
  2012MNRAS.424.1762M,2018ApJ...854..162S} have found that the widths
of the core and the conal components have similar distributions with
pulsar period and identical lower boundaries. This results confirmed
that the core and the conal components originate from similar heights.
As a result in the current understanding of the emission beam
structure the core component is associated with the central field
lines around the dipolar magnetic axis, the inner cones the
intermediate field lines between the axis and the open field line
boundary, while the outer cones occupy the outermost field lines
bordering the boundary.  It is expected that the radius of curvature
of the magnetic field lines is highest in the core region and
progressively decreases towards the inner and outer cones.

Alternative models for the emission beam include the ''patchy cone''
\citep{1988MNRAS.234..477L} where the components are expected to arise
from similar heights but are distributed randomly within the emission
beam. The primary justification for this model comes from the presence
of the ``partially conal'' profiles where the PPAs are asymmetric and
the SG point is shifted towards one edge of the profile. Later studies
have shown that in most of these cases the beam shape follows a
core-cone structure with one edge of the profile having lower
intensity most of the time and only appearing during flaring events in
the single pulses \citep{2011ApJ...727...92M}. In a few other cases
additional components in the form of pre/post cursor emission are seen
that are not part of the emission beam \citep{2015ApJ...798..105B}.
There are also suggestions that the emission beam being a ``fan beam''
with the components arising from emission along flux tubes
\citep{1987ApJ...322..822M,2012MNRAS.420.3403D,2014ApJ...789...73W,
  2015MNRAS.446.2505D}. The fan beam has been used to explain the
bifurcated features seen in the profile of certain pulsars. However,
its not clear if the characteristics of the different profile
morphologies can be explained in this model. Further, the requirement
of a tightly packed sparking pattern above the polar cap, that gives
rise to subpulse drifting (see discussion below) is also difficult to
reconcile with a fan beam without conal symmetry.

There are indications that the emission beam shows evolution with the
spin-down energy loss ($\dot{E}$) or characteristic age (see e.g.
\citep{2007MNRAS.380.1678K}).  The profile classes are not distributed
uniformly with $\dot{E}$, with S$_t$ more prevalent in the high
$\dot{E}$ range above 10$^{33}$ erg~s$^{-1}$, while the conal profiles
corresponding to D and S$_d$ types are usually seen in pulsars with
low $\dot{E}<10^{32}$ erg~s$^{-1}$
\citep{2019MNRAS.482.3757B,2022MNRAS.514.3202R}.  The differences in
beam shape presented in \citep{2008MNRAS.391.1210W} are subtle, with
the fractional polarization between the two subsets being more
prominent.  Other possibilities include the increase of the number of
sparks in high $\dot{E}$ pulsars with wider open field line region,
which form a large number of closely packed components with blurred
boundaries between them \citep{2000ApJ...541..351G}.

\subsection{Measuring Spectral Variation across the Emission Beam}
The coherent radio emission from pulsars have a steep power law
spectra between 100 MHz and 10 GHz with spectral index $\sim-1.8$
\citep{2000A&AS..147..195M, 2018MNRAS.473.4436J}. The emission beam
studies have shown that the central core component in the profiles are
associated with the field lines close to the magnetic axis where the
radius of curvature is larger, while the conal emission occupies the
field lines closer to the boundary of the open field line regions
where the magnetic field lines are more curved with relatively smaller
values of the radius of curvature. There have been indications that
the core components have steeper spectra compared to the cones, e.g.
the appearance of conal outriders at higher frequencies in S$_t$
profile types. However, detailed measurements of the relative spectral
indices of the components have not been carried out due to several
challenges in estimating the flux density levels of the profile. This
primarily arises from the lack of proper instrumental calibration
required to scale the measured signal to the appropriate flux
densities. In addition, the pulsar signal is also affected by
scintillation over timescales varying between several minutes
(diffractive scintillation) and extending to several months
(refractive scintillation) \citep{1990ARA&A..28..561R}. As a result
the proper estimation of the flux densities require averaging over
long timescales which is only available for relatively few pulsars at
specific frequencies \citep{1977AJ.....82..701H,
  1992ApJ...392..530K,1994AJ....108.1854L,2000ApJ...539..300S,
  2021MNRAS.501.4490K}. Other issues affecting flux density estimates,
particularly at lower frequencies, involve interstellar scattering
\citep{2015MNRAS.449.1570L}. One attempt to measure the spectral index
of the different components in the profile by
\cite{1993A&A...272..268G} used the Gaussian fitting technique to
identify the components. No clear trend of spectral variation between
components of the profile was found in this work. However, the
Gaussian components were often displaced from actual profile
components and made it difficult to have a direct connection between
them.

A way around the issue of non-availability of proper flux calibration
of the profiles as well as variations due to interstellar
scintillation has been devised in our previous works
\cite{2022ApJ...927..208B}, where it was noticed that these effects
introduce an unknown but identical multiplicative scaling for all
components in the profile. As a result the ratio between the component
intensities are expected to be invariant of the measurement conditions
and can be used to estimate the relative spectral index of the core
with the inner and outer conal components as follows :
\begin{equation}
S_{core} = S_1~ \nu^{\alpha_{core}};~~~
S_{cone} = S_2~ \nu^{\alpha_{cone}};~~~
S_{in} = S_3~ \nu^{\alpha_{in}};~~~
S_{out} = S_4~ \nu^{\alpha_{out}} \nonumber.
\end{equation}
\begin{eqnarray}
\left(\frac{S_{core}}{S_{cone}}\right) & = & \left(\frac{S_1}{S_2}\right) \nu^{\Delta\alpha_{core/cone}} ~~~~~ \Delta\alpha_{core/cone} = \alpha_{core} - \alpha_{cone} \nonumber \\
\left(\frac{S_{core}}{S_{in}}\right) & = & \left(\frac{S_1}{S_3}\right) \nu^{\Delta\alpha_{core/in}} ~~~~~ \Delta\alpha_{core/in} = \alpha_{core} - \alpha_{in} \nonumber \\
\left(\frac{S_{core}}{S_{out}}\right) & = & \left(\frac{S_1}{S_4}\right) \nu^{\Delta\alpha_{core/out}} ~~~~~ \Delta\alpha_{core/out} = \alpha_{core} - \alpha_{out} \nonumber \\
\left(\frac{S_{in}}{S_{out}}\right) & = & \left(\frac{S_3}{S_4}\right) \nu^{\Delta\alpha_{in/out}} ~~~~~ \Delta\alpha_{in/out} = \alpha_{in} - \alpha_{out} \nonumber \\
\end{eqnarray}
The estimations of the relative spectral index
($\Delta\alpha_{core/cone}$) of the core and the conal components are
possible in pulsar profiles with central LOS traverses, such that both
the core and conal components are present, i.e. T, T$_{1/2}$ and M
type profiles. In addition, the relative spectral index between the
inner and the outer conal pairs ($\Delta\alpha_{in/out}$) can be
carried out in M, $_c$Q and $_c$T profiles.

\begin{figure}[H]
\includegraphics[width=13.0cm]{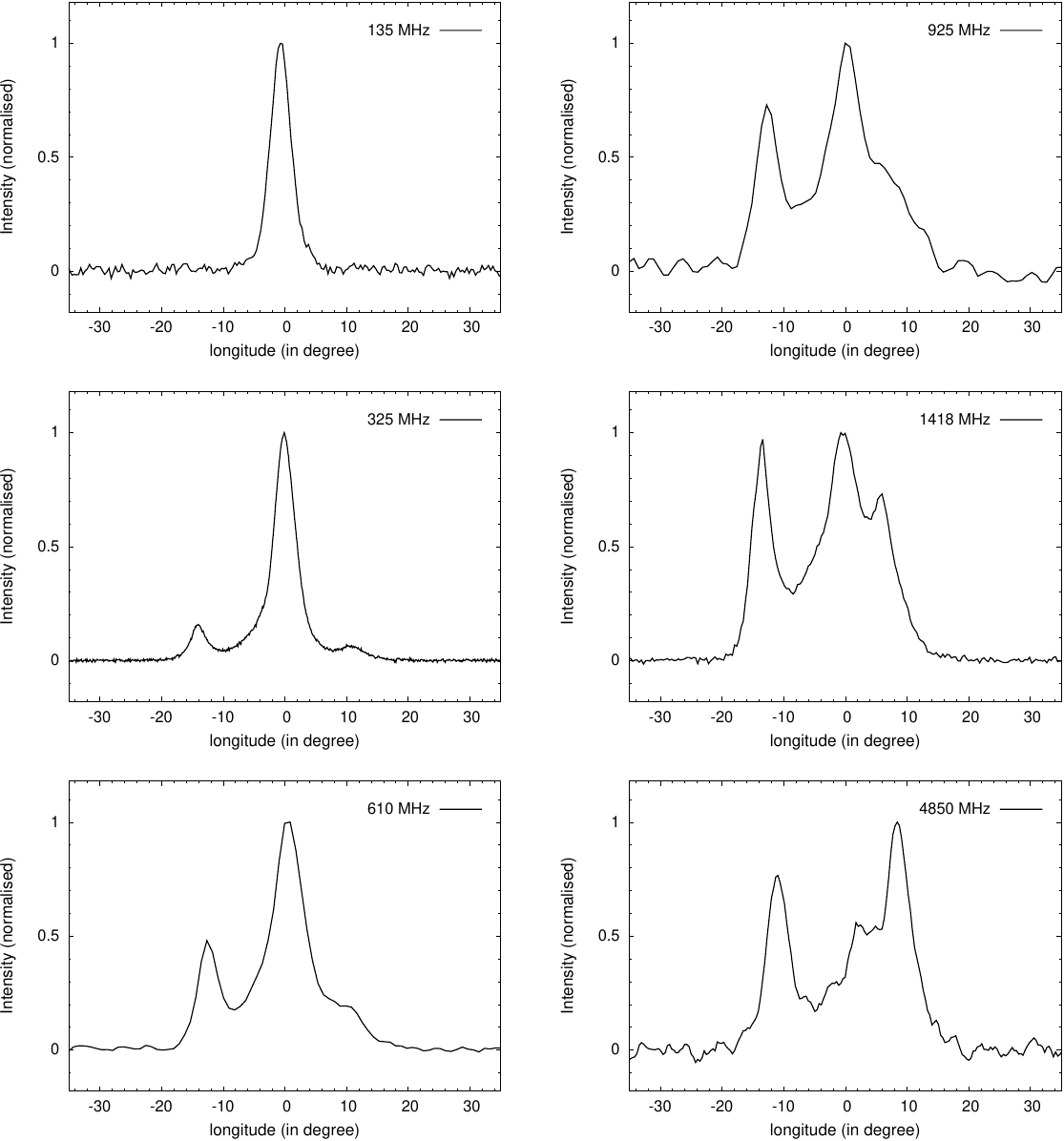}
\caption{The Figure shows the T type average profile of PSR B1821+05 at 6 
different frequencies at 135 MHz (top left), 325 MHz (middle left), 610 MHz 
(bottom left), 925 MHz (top right), 1408 MHz (middle right) and 4850 MHz 
(bottom right). The core component in the center of the profile is stronger at
the lower frequencies and becomes weaker compared to the surrounding cones with
increasing frequency.}
\label{fig:profB1821}
\end{figure}

\begin{figure}[H]
\includegraphics[width=10.0cm]{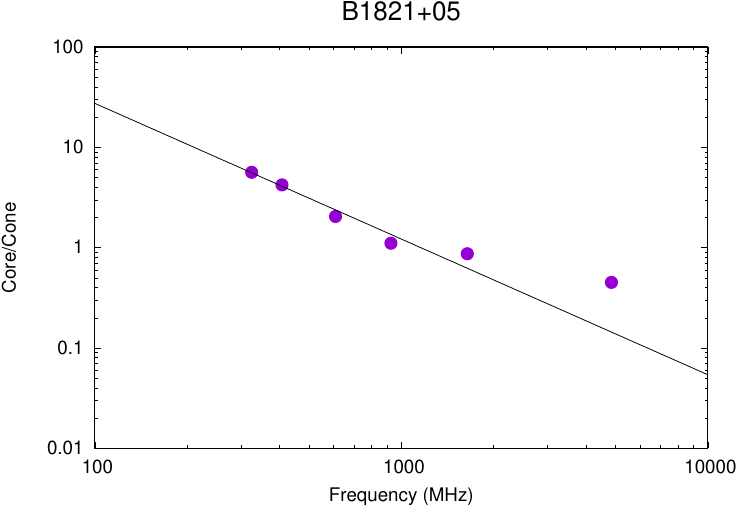}
\caption{The Figure shows the frequency evolution of the relative intensities 
of the core and the conal components ($S_{core}/S_{cone}$) of PSR B1821+05 
between 100 MHz and 10 GHz. {\bf The power law fit to the relative spectral 
	index shown by solid black line has spectral index of -1.55$\pm$0.1.} }
\label{fig:specB1821}
\end{figure}

Fig.~\ref{fig:profB1821} shows frequency evolution of the profile
shape of PSR B1821+05 that has a T type profile. The figure shows the
measured profiles at six frequency bands between 100 MHz and 5 GHz. At
the lowest frequency of 135 MHz only the core component is seen, while
the conal components are too weak to be detected. With increasing
frequency the relative intensity of the conal components increases
with respect to the core and the cones dominate the profile at
frequencies above 1 GHz. Fig. \ref{fig:specB1821} shows the spectral
plot of the relative intensities between the core and the cone
($S_{core}/S_{cone}$) between 100 MHz and 10 GHz, with the estimated
spectral index $\Delta\alpha_{core/cone} = -1.55\pm0.10$ (see line in
Fig.  \ref{fig:specB1821}), demonstrating the core to have a much
steeper spectral index compared to the conal emission. At the lowest
frequency range the conal components are not visible in the profile
and hence the relative intensities cannot be measured. The core
emission at 4.8 GHz merges with the trailing conal component (see Fig.
\ref{fig:profB1821}, bottom right panel), leading to imprecise
measurements. As a result at the highest frequency the relative
intensities deviate from the linear spectral behaviour.

Similar studies have been carried out in around 50 pulsars having both
core and conal components in their profiles
\citep{2022ApJ...927..208B} and in all cases the core component have a
steeper spectra than the cones, with $\Delta\alpha_{core/cone}$
varying between -0.4 and -2.0. The radius of curvature of the magnetic
field lines are expected to change between the core and the conal
components and the only known emission mechanism that has direct
dependence on the radius of curvature of the field lines is CCR from
charge bunches. As a result the observational result of the core
having a steeper spectral index than the conal components within the
pulsar profile strongly favours the radio emission mechanism in
pulsars to be CCR.

\section{Subpulse Drifting : Window into Inner Acceleration Region}
\label{sec4}
The CCR from charge bunches needs the presence of a multi-component
electron-positron plasma in the form of plasma clouds, separated by
less dense regions filled with positively charged ions, and requires a
non-stationary plasma flow. In this section we briefly summarize our
understanding of the plasma generation process in pulsars and its
observational signature in the form of subpulse drifting.

The plasma is expected to be generated in an inner acceleration region
(IAR) above the polar cap extending to heights of around 10-100 m. In
one of the earlier studies by \cite{1975ApJ...196...51R} the IAR was
considered to be a Vacuum Gap (IVG). The IVG is characterised by very
high magnetic fields ($\sim10^{13}$ G) as well as large electric
potential difference (10$^{12}$ V) along the magnetic field. Under
these extreme conditions regular breakdown of the gap in the form of
spark discharges is expected. This mechanism of plasma generation is
possible in pulsars where $\vec{\Omega}\cdot\vec{B} < 0$ (here
$\Omega=2\pi/P$ is the angular velocity of the neutron star), such
that positively charged particles are accelerated away from the
stellar surface while the negatively charged particles are accelerated
towards the surface. In the IVG the electron-positron pairs are
produced from ambient $\gamma$-ray photons and accelerated to high
energies with Lorentz factor $\gamma_b\sim10^6$
\citep{1971ApJ...164..529S}. The positrons are accelerated away from
the gap and produce additional $\gamma$-ray photons due to curvature
radiation or inverse compton scattering as they propagate along the
curved magnetic field lines.  These additional $\gamma$-ray photons
generate further pairs resulting in a cascading effect. Effective
break down of the gap requires the magnetic field to be highly curved
and non-dipolar in nature such that high energy $\gamma$-ray photons
can be produced. In this process a primary plasma beam comprising of
high energetic positrons is formed.

When the charge density in the IVG equals the Golreich-Julian density,
$\rho_{GJ} = \Omega B/2\pi c$ \citep{1969ApJ...157..869G}, the
electric potential difference along the magnetic field is fully
screened and pair production is halted. The potential difference
reappears once the plasma escapes the gap due to inertial motion and
the next sparking event is triggered forming the non-stationary plasma
flow. The primary particles continue to produce additional pairs
outside the IAR to form clouds of secondary pair plasma that are
considerably less energetic, with Lorentz factors, $\gamma^{sp}_s$
between 100 and 1000, and has high multiplicity, $\kappa^{sp}\sim10^4$
\citep{1971ApJ...164..529S}.  The CCR develops in these spark
associated secondary plasma clouds that stream outwards along the open
magnetic field line region.

Subsequent works have revealed several limitations of the IVG model of
IAR. The back-flowing electrons produced during sparking are expected
to heat the surface to temperatures of $10^6$ K, that are near the
critical level for ions ($T_i$) to be emitted freely from the surface
and screen the potential difference along the magnetic field
\citep{CR80,1986MNRAS.218..477J}. In a purely vacuum gap there is no
mechanism to constrain the lateral size of the sparks as there is
unscreened potential difference in the boundary between two adjacent
sparks. As a result the primary particles cannot be confined to any
specific location within the IVG, but likely scatter in the direction
opposite to principal normal to the curvature of the local magnetic
field lines \citep{CR80}.

In order to address the heating of the surface due to back-flowing
electrons during sparking an updated model of the IAR has been
proposed. It has been suggested that the IAR is not a complete vacuum
but has a steady supply of positively charged ions from the surface
and forms a Partially Screened Gap (PSG) \citep{2003A&A...407..315G}.
The potential difference along the gap is reduced by a screening
factor $\eta = 1 - \rho_i/\rho_{GJ}$, where $\rho_i$ is the average
density of ions in the gap. In a PSG the sparking process is primarily
a mechanism to regulate the surface temperature around $T_i$ of ionic
free flow. When the surface temperature ($T_S$) is close to $T_i$
there is free flow of ions from the surface and the potential
difference along the IAR is screened. As the surface cools down and
$T_S < T_i$, the supply of ions from the surface decreases such that
the density goes below $\rho_{GJ}$ and the potential difference
appears in the IAR. This starts the sparking process and the primary
plasma is produced along with the back-flowing electrons that once
again heat the surface to critical level thereby terminating the
spark. The typical timescale for each spark to develop and empty the
IAR is around 30 $\mu$seconds, while the surface cools rapidly once
the sparking stops with typical cooling time of 30 nseconds.

When a spark starts at a location within IAR, it also spreads out
across the field lines till enough particles are produced to heat the
surface and screen the potential difference along the magnetic field.
As a result the lateral size of the sparks ($h_\perp$) is regulated by
the energy that needs to be deposited on the surface for the thermal
regulation process. The PSG is the only known mechanism for confining
sparks in the IAR, where the sparking location is determined by the
surface temperature profile. The sparks are formed in a tightly packed
manner for effective thermal regulation of the surface. The peak
density is near the center of the spark and the density gradually
decreases towards the edge, with the boundary between two sparks
dominated by the ions emitted from the surface.  The primary plasma
produced in the sparks leave the IAR and they give rise to the columns
of secondary plasma clouds. The inter-cloud regions correspond to the
boundary between sparks, and are dominated by the positively charged
ions. For a typical normal pulsar the Lorentz factor of the secondary
pair plasma is $\gamma^{sp}_s = \gamma^{isp}_s \sim 100$ and the
Lorentz factor of the ion component $\gamma^{sp}_{ion} \sim 10^3$,
where superscript ``sp'' and ``isp'' stands for the spark and
interspark region. The multiplicity $\kappa^{sp} \sim 10^4$ at the
center of the spark which gradually becomes $\kappa^{isp} \sim 1$ in
the interspark region. The Lorentz factor of the ion in the interspark
region is $\gamma^{isp}_{ion} \sim 10$. In terms of pulsar parameters
$P$ and $\dot{P}_{15}$ and $\mathcal{R}$, the plasma frequency of the
pair plasma is $\omega_p = 6.4151\times10^{4}~\kappa^{0.5}
\mathcal{R}^{-1.5}P^{-1.75} \dot{P}_{15}^{0.25}$ and the ion plasma
frequency is $\omega _{p,ion} =1.02\times 10^{3} ~\mathcal{R}^{-1.5}
P^{-1.75} {\dot{P}_{-15}}^{0.25}$, where ion $^{56} Fe_{26}$ species
is used (see \cite{2023ApJ...952..151M} for these estimates).

When the pulsar magnetosphere is filled up with $\rho_{GJ}$, the
plasma co-rotates with the star due to $\vec{E}\times\vec{B}$ drift.
During sparking the plasma density in the IAR is less than $\rho_{GJ}$
and plasma no longer co-rotates with the star. The sparks during their
lifetimes lag behind the rotation of the star
\citep{2023ApJ...947...86B}, and the location of maximum
heating/cooling shifts opposite to the direction of rotation as the
spark develops. Once the plasma empties and the surface cools down the
next spark is formed slightly shifted in the direction opposite to
rotation.  This lack of co-rotation is imprinted in the plasma clouds
as they move along the open field line region and emit radio waves.
The effect is seen in the single pulse behaviour of certain pulsars as
the phenomenon of subpulse drifting
\citep{1973ApJ...182..245B,2006A&A...445..243W,2016ApJ...833...29B}.
The subpulse drifting is intricately connected to the conditions in
the IAR, like the potential difference along the magnetic field, the
nature of non-dipolar magnetic fields, the surface conditions, etc. We
describe below the measurement of the drifting behaviour in pulsars
and how these results can be used to obtain observational evidence for
the presence of non-dipolar fields above the stellar surface and for
the IAR to be a PSG. This in turn puts the presence of plasma clouds
in the open field line region, the presence of multi-component plasma,
and the dominance of $^{56} Fe_{26}$ ions between plasma clouds, on a
firm observational footing.

\begin{figure}[H]
	\begin{center}
\includegraphics[width=5.0cm]{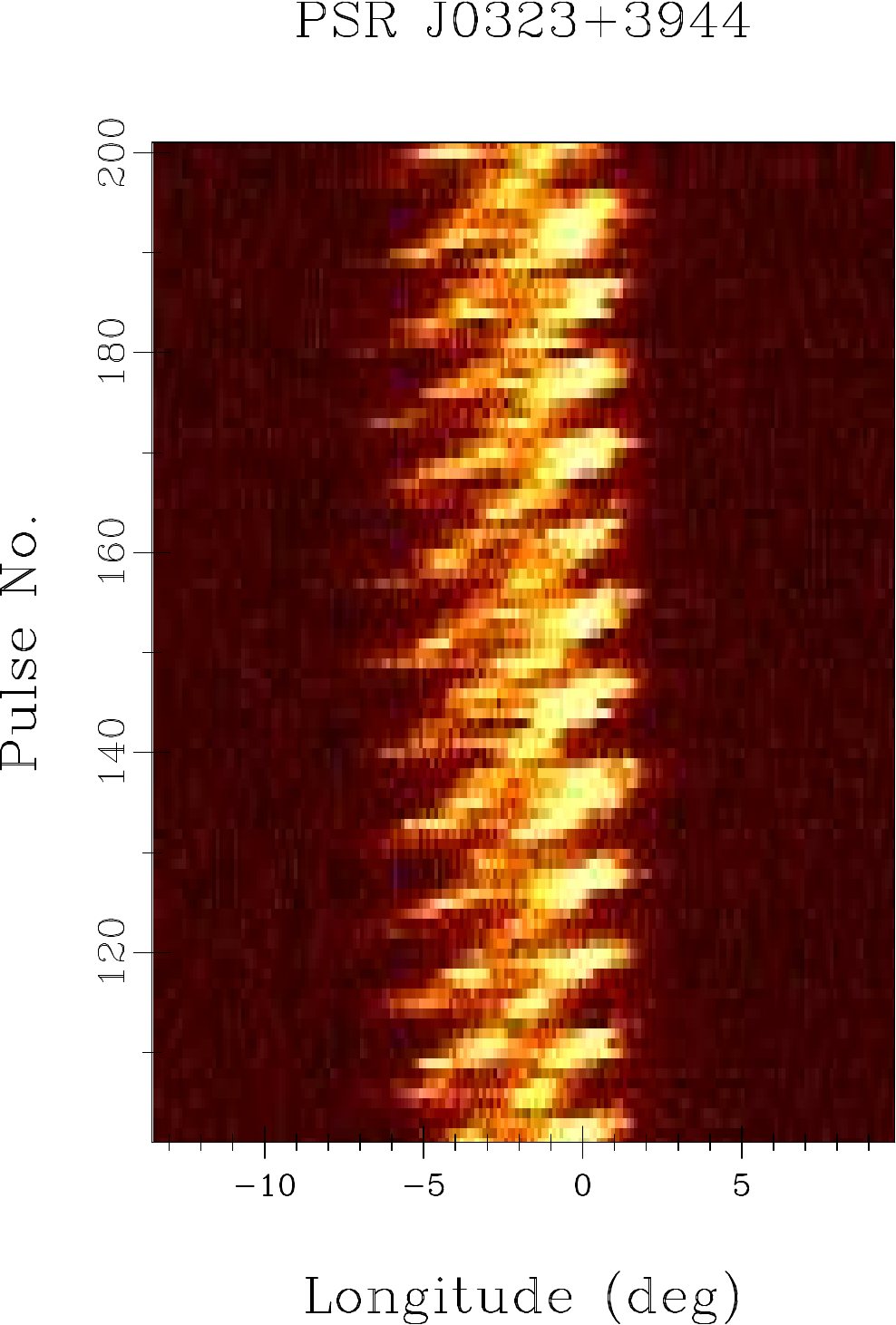}
\includegraphics[width=5.0cm]{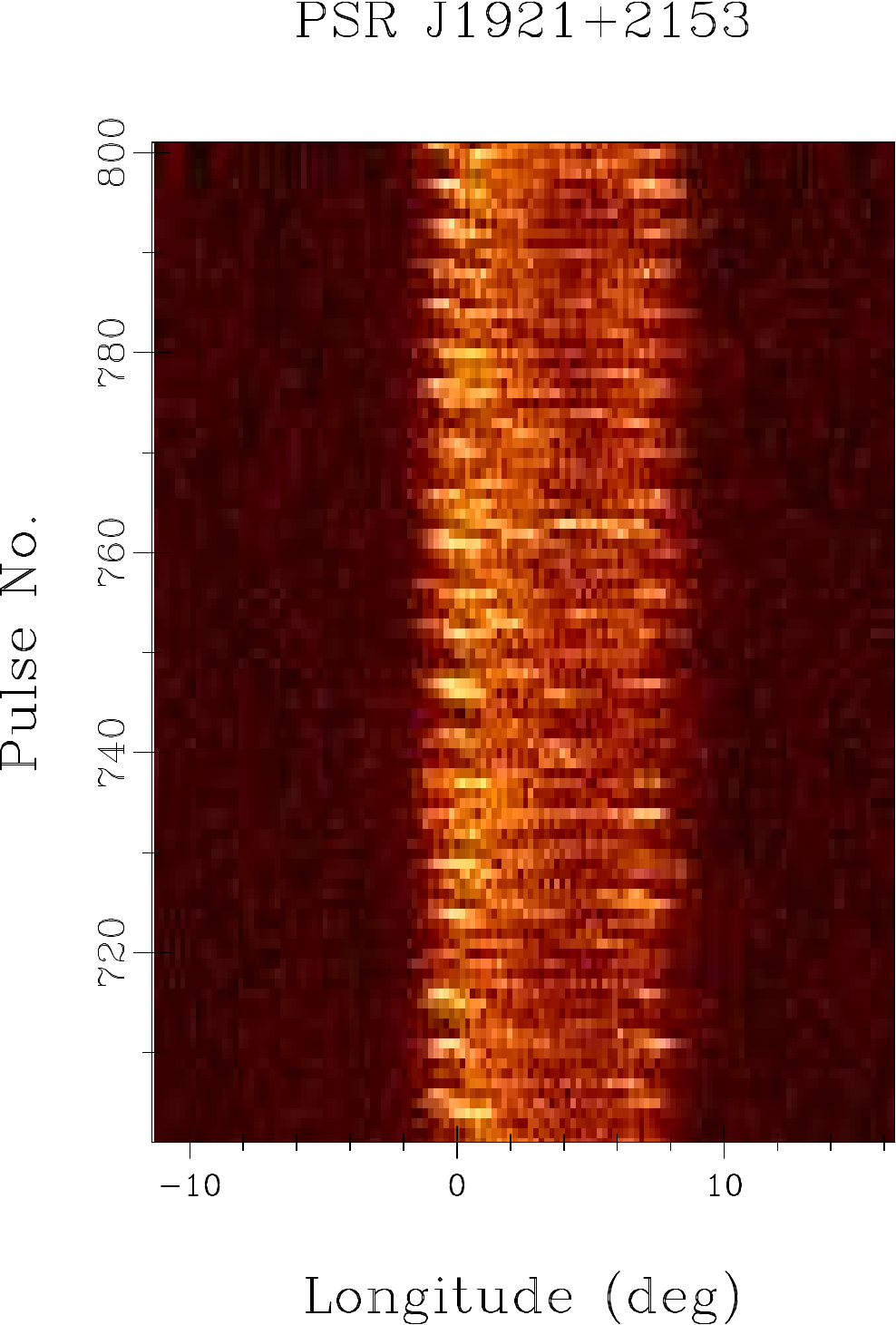}
	\end{center}
        \caption{The Figure shows single pulse stack of PSR J0323+3944
          (or PSR B0320+39) at 333 MHz (left panel) and PSR J1921+2153
          (or PSR B1919+21) at 610 MHz (right panel), where the
          subpulse drifting phenomenon is clearly observed. The data
          is obtained from the GMRT and for details see
          \cite{2019MNRAS.482.3757B}.}
      \label{fig:driftexample}
\end{figure}
\subsection{Drift Phase Variations : Evidence of Non-dipolar Magnetic Field}
\begin{figure}[H]
\includegraphics[width=13.5cm]{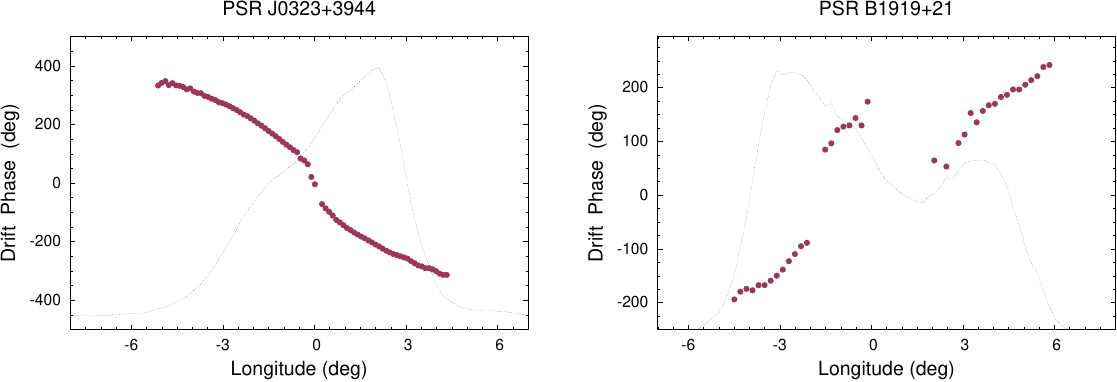}
\caption{The Figure shows the phase variation of the drifting
  subpulses across the profile of PSR J0323+3944 at 333 MHz (left
  panel) and PSR B1919+21 at 610 MHz (right panel).}
\label{fig:phslrfs}
\end{figure}

The single pulse emission in pulsars comprise of one or more
constituents that are called subpulses. In certain cases the subpulses
show systematic periodic shifts and the phenomenon is called subpulse
drifting which was first discovered by \cite{1968Natur.220..231D}. The
drifting behaviour is best visualized when the single pulses are
represented in the form of a pulse stack which is a two dimensional
plot with the rotation longitude along the x-axis and the pulse number
in the y-axis. Two examples of subpulse drifting is shown in
Fig.~\ref{fig:driftexample}. The nature of the drifting behaviour is
associated with the profile morphology and LOS geometry
\citep{2019MNRAS.482.3757B}. The pulsars with large shift of the
subpulses across the emission window have conal profiles of S$_d$ and
$D$ types (see previous section). On the contrary the M type profile
usually have very little shift in position of the subpulses but
periodic modulation of intensity in their conal components. The
central core component in these profiles do not show any drift
behaviour. The intermediate LOS profiles like $_c$T and $_c$Q show
complex drift behaviour with reversals in drift direction, also known
as bi-drifting, between different components of the profile in a few
cases. The drifting behaviour is measured using the fourier
transformation technique, like the Longitude Resolve Fluctuation
Spectra (LRFS) \citep{1975ApJ...197..481B}, where Fourier transforms
are carried out along each longitude range of the pulse stack for a
fixed number of pulses. The drifting periodicity ($P_3$), that
represents the time interval between subpulses repeating at any
longitude, is seen as a peak frequency in the LRFS. The phase
variations corresponding to the peak frequency shows the track of the
subpulses across the profile window (see Fig.~\ref{fig:phslrfs}).

As discussed earlier the sparks in a PSG are formed in tightly packed
configuration and lag behind the rotation motion of the star
\citep{2023ApJ...947...86B}. The subsequent sparks are formed slightly
shifted in the direction opposite to the rotation of the star. In
addition, the well defined polar cap boundary separating the open and
closed field line region of the magnetosphere constrains the later
sparks to be formed along the boundary.  Hence, the two dimensional
sparking pattern in the IAR evolves with time along two different
directions, in a clockwise and counterclockwise manner in the two
halves of the polar cap. In the center of the IAR the sparks are
formed at regular intervals in roughly the same location since there
is no space for shifting. The central sparks makes up the core
component in the profile that does not show any drifting behaviour.

\begin{figure}[H]
\includegraphics[width=13.5cm]{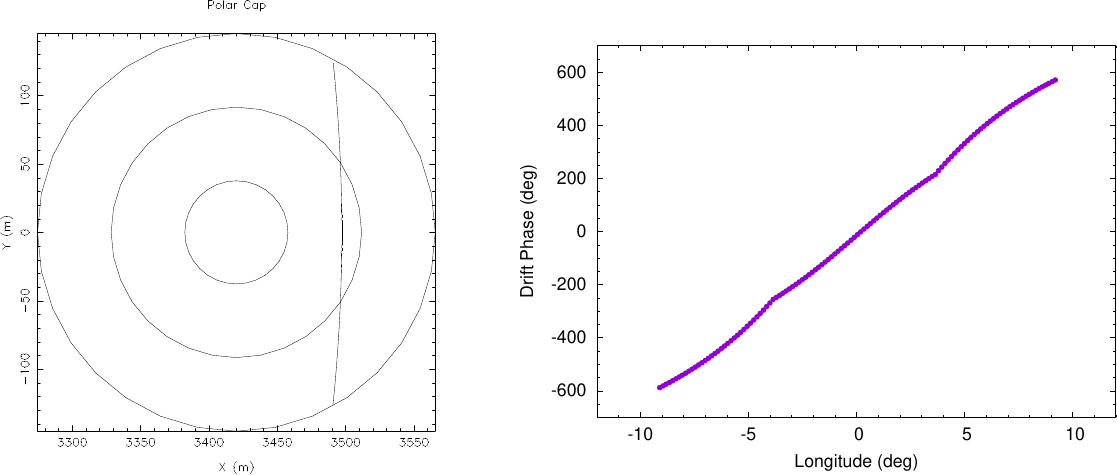}
\caption{The Figure shows the modelled phase variations of subpulse drifting
from a polar cap with dipolar magnetic field. The left panel shows the LOS 
traverse across the polar cap, where the three concentric rings demarcate the
core, inner and outer conal regions. The LOS traverse cuts across the polar cap
about halfway between the axis and the open field line boundary. The right 
panel shows the simulated phase variations of the drifting subpulses for this
configuration.}
\label{fig:phsDip}
\end{figure}

The drifting periodicity as well as the phase variations can be
associated with the sparking process in the IAR. The drifting
periodicity measures the electric potential difference and
particularly the screening factor, $\eta$, of PSG and will be
discussed in the next subsection. The phase variations across the
profile traces the evolutionary track of the sparking pattern along
the projection of the LOS (see Fig.~\ref{fig:phsDip}, left panel). As
a result the phase behaviour has information regarding the shape of
the polar cap which is determined by the nature of the surface
magnetic field. In case of purely dipolar field the phase variations
are expected to be mostly linear with small deviations near the edge
of the inner and outer cones. The expected phase variations of
drifting for an intermediate LOS cut located midway between the
magnetic axis and the polar cap edge has been simulated and shown in
Fig.~\ref{fig:phsDip} (see \cite{2023ApJ...947...86B} for details).
The phase behaviour is mostly linear across the entire profile with
small deviations at the crossing between the inner and the outer
cones, and near the edges. The measured phase variations for two
pulsars PSR J0323+3944 (left panel) and PSR B1919+21 (right panel) is
shown in Fig.~\ref{fig:phslrfs}. The phases show larger deviation from
linearity and significant jumps between components, confirming that
the surface magnetic fields in pulsars are non-dipolar in nature. The
direction of the slope of the phase changes are reversed in these two
pulsars and has a degenerate origin, either due to the drifting
periodicity being aliased or the local magnetic field on the surface
being highly twisted \citep{2023ApJ...947...86B}.
\subsection{Dependence of Drift Periodicity with $\dot{E}$ : Evidence of 
Partially Screened Gap}
The drifting periodicity is the time taken by subsequent sparks to lag
behind the extent of the spark diameter, such that the subpulse peaks
appear at the same longitude in the emission window. The speed at
which the sparking pattern shifts in the gap is $v_{sp} = \eta(E/B)c$,
and the repetition time is estimated as $P_3 =h_{\perp}/v_{sp}=
1/2\pi\eta\cos{\alpha_l}$ \citep{2023ApJ...947...86B}, here $\alpha_l$
is the angle of the local magnetic field with the rotation axis.

The total energy outflow from the PSG ($L_{PSG}$) due to the
outflowing plasma can be estimated as \citep{2016ApJ...833...29B} :
\begin{equation}
L_{PSG}\simeq\gamma_b m_{e}c^3\eta n_{GJ}A_{pc}.
\label{L_PSG}
\end{equation}
Here, $n_{GJ} = \rho_{GJ}/e$, and $A_{pc}$ is the area of the polar
cap. The quantity $n_{GJ}A_{pc}$ is invariant of the nature of the
surface fields due to conservation of magnetic flux and can be
estimated as
\begin{equation}
n_{GJ}A_{pc} = 2\times10^{19} \cos{\alpha_l} (\dot{P}_{-15}/P^3)^{0.5},
\end{equation}
Additionally, $\dot{E}$ is also obtained from $P$ and $\dot{P}_{-15}$
in the form $\dot{E} = 4\times10^{31} (\dot{P}_{-15}/P^3)$
erg~s$^{-1}$. The ratio between $L_{PSG}$ and $\dot{E}$ has the form :
\begin{equation}
\xi = \frac{L_{PSG}}{\dot{E}} \simeq 1.2\times10^{-8} \eta \gamma_b \cos{\alpha_l}\left(\frac{\dot{E}}{\dot{E}_1}\right)^{-0.5},
\end{equation}
Here, $\dot{E}_1 = 4\times10^{31}$ erg~s$^{-1}$. The PSG model
provides a direct connection between two independent measurable from
pulsars, $P_3$ and $\dot{E}$ which has the form
\begin{equation}
P_3\simeq 2\times10^{-9}\left(\frac{\gamma_p}{\xi}\right)\left(\frac{\dot{E}}{\dot{E}_1}\right)^{-0.5},
\end{equation}
i.e. we obtain a dependence $P_3\propto\dot{E}^{-0.5}$ between the two 
quantities. 

\begin{figure}[H]
\includegraphics[width=10.0cm]{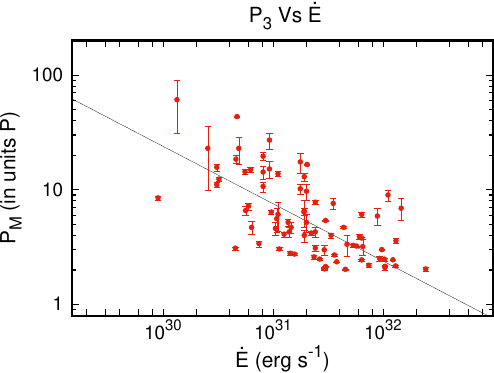}
\caption{The Figure shows the distribution of drifting periodicity ($P_3$) with
the spin-down energy loss ($\dot{E}$). The black line shows the expected $P_3 
\propto \dot{E}^{-0.5}$ dependence from the Partially Screened Gap model of the
Inner Acceleration Region.}
\label{fig:P3Edot}
\end{figure}

The distribution of $P_3$ with $\dot{E}$ has been estimated in a
number of works
\citep{2016ApJ...833...29B,2019MNRAS.482.3757B,2020ApJ...889..133B}
and reproduced in Fig.~\ref{fig:P3Edot}, with additional $P_3$
measurements from recent works added to the figure
\citep{2019MNRAS.486.5216B,
  2020MNRAS.499..906B,2021MNRAS.500.4139R,2022ApJ...929...71W,
  2022ApJ...934...23S,2023ApJ...950..166W,2023MNRAS.526..691B,
  2023MNRAS.526.3184Y}. The drifting behaviour is often conflated with
the periodic nulling and periodic modulation features seen in pulsars
\citep{2023MNRAS.520.4562S}, that have quasi-periodic variation
unrelated to the sparking process in the IAR
\citep{2020ApJ...889..133B}. We have restricted the $P_3$ in
Fig.~\ref{fig:P3Edot} to cases with clear indications of drifting in
the single pulses. Although, there is a prospect of aliasing
associated with the $P_3$ measurements, where any intrinsic $P_3 < 2P$
is seen as a higher periodic feature $P_3^a = 1/(1-1/P_3) > 2P$, the
general trend of the anti-correlation between $P_3$ and $\dot{E}$ is
evident in Fig.~\ref{fig:P3Edot}. Additional support for the presence
of such dependence is also seen in the grouping of periodicities at
the alias boundary $2P$ near $\dot{E}=10^{32}$ erg~s$^{-1}$. Some
pulsars above this range shows an increase in their $P_3$ suggesting
an aliased measurement, which will be expected from the
anti-correlation. The figure also shows the $\dot{E}^{-0.5}$
dependence (black line) of the PSG model which is consistent with the
distribution. The scatter in the plot is also explained in this model
due to the dependence on parameters like $\gamma_p$ and $\xi$, that
are likely to vary between different pulsars. The presence of
anti-correlation between two independent measurables, $P_3$ and
$\dot{E}$, provide strong observational justification for the IAR to
be a PSG.


\section{Summary}
The identification of highly linearly polarized signals following the
RVM in conjunction with the findings that pulsar radio emission from
normal pulsars detach from regions below 0.1$\mathcal{R}$ strongly
suggest that CCR is the radio emission mechanism in pulsars. The OPMs
indicate that the linearly polarized orthogonal modes are excited in
strongly magnetized pair plasma that can reach the observer as X- and
O-mode. The observed circular polarization suggests the presence of an
ion component in the plasma. The multiple-component average profiles,
the single pulse quasi-periodic structure and subpulse drifting
requires non-stationary plasma flow and strong non-dipolar surface
magnetic fields. Hence, these observational evidences provide the
framework of the necessary conditions for CCR to develop.

CCR from charge bunches has been considered as a plausible candidate for 
radio emission mechanism from the beginning of pulsar research. However, there are still some questions raised regarding a suitable theory that can justify the existence of a stable charge bunches in pulsar plasma. Alternate emission mechanisms such
as maser mechanism has also been suggested for the radio emission in pulsars, however these theories do not have adequate observational support.
Studies addressing the issue of charge bunch formation on the other hand 
resulted in identifying the existence of stable relativistic charged solitons in the 
pulsar plasma \citep{2000ApJ...544.1081M,2018MNRAS.480.4526L,2022MNRAS.516.3715R}.
These studies are performed in the one dimensional approximation, whereas realistically
stable charge bunch formation needs to be established in two or three dimensions.
The overwhelming observational evidence for CCR in pulsars strongly motivates
theoretical research in these directions. In the accompanying article by 
Melikidze, Mitra \& Basu the development and current state of the theory of 
charge bunch formation in pulsar plasma is reviewed. 

The enhanced sensitivity of radio Telescopes in conjunction with
improved data analysis methods in recent years have revealed a
plethora of interesting new phenomena from pulsars.  While we are
fairly confident that the radio emission mechanism from normal pulsars
is CCR, there are other issues that still remain unclear. Below we
list some of these key problems associated with the pulsar activity.
\begin{enumerate}
\item The origin and shape of the pulsar emission beam and its
  evolution with $\dot{E}$ (see e.g. \cite{2007MNRAS.380.1678K}).
	\item The physical origin of the evolution of average profile
          width, component width and component separation with
          frequency, and their dependence on the emission height (see
          accompanying paper Melikidze, Mitra \& Basu, also see e.g.
          \citep{2002ApJ...577..322M}).
	\item The origin of inter-pulse emission and the
          post/pre-cursor and off-pulse emission outside the main
          pulse window (e.g.
          \cite{2019MNRAS.490.4565J,2015ApJ...798..105B,2020ApJ...905...30B,2023ApJ...949..115Y}).
	\item The complete absence of subpulse drifting in pulsars
          with $\dot{E} > 5\times10^{32}$ erg~s$^{-1}$ and also in
          around 50\% of the population below this limit (e.g.
          \cite{2006A&A...445..243W,2019MNRAS.482.3757B,2023MNRAS.520.4562S}).
	\item What causes the phenomenon of mode (or state) changing and to explain the observed correlation/anticorrelation between radio and X-ray intensities during mode changing (\cite{2018MNRAS.480.3655H}. Also what causes the weak emission detected in the null state of pulsars (\cite{2023NatAs...7.1235C,2024arXiv240301084Y}).
	\item How does pulsar signals escape the pulsar magnetosphere? (see e.g. \cite{2014ApJ...794..105M,2022MNRAS.512.3589R})
        \item It is worth mentioning that, unlike normal pulsars, in
          millisecond pulsars the location of the radio emission
          region has not been constrained and as a result the coherent
          radio emission mechanism is still unknown (see e.g.
          \cite{1998ApJ...501..270K,1999ApJ...526..957K,1998ApJ...501..286X})
          ?
	\item What is the origin of radio emission from Magnetars (see
          e.g. \cite{2015RPPh...78k6901T})?
\end{enumerate}
The above list is not exhaustive but gives a flavour of the exciting
scientific problems that can be explored in the future.

%
%
\vspace{6pt}

\funding{D.M. acknowledges the support of the Department of Atomic
Energy, Government of India, under project No. 12-R\&DTFR-5.02-0700.
This work was supported by the grant 2020/37/B/ST9/02215 of the National Science Centre, Poland.}

\acknowledgments{We thank the
staff of the GMRT, who made these observations possible.
D.M. acknowledges the support of the Department of Atomic
Energy, Government of India, under project No. 12-R\&DTFR-5.02-0700.
This work was supported by the grant 2020/37/B/ST9/02215 of the National Science Centre, Poland.}

\conflictsofinterest{The authors declare no conflict of interest.} 



\abbreviations{Abbreviations}{
The following abbreviations are used in this manuscript:\\

\noindent 
\begin{tabular}{@{}ll}

        CCR & Coherent Curvature Radiation\\
        GMRT & Giant Metrewave Radio Telescope \\
        IAR & Inner Acceleration Region\\
        IVG  & Inner Vacuum Gap \\
        LOS & Line Of Sight \\
        LRFS & Longitude Resolve Fluctuation Spectra\\
        $lt$-mode & Longitudinal Transverse mode\\
        OPM & Orthogonal Polarization Mode \\
        O-mode & Ordinary mode \\
        PPA & Polarization Position Angle \\
        PSG & Partially Screened Gap\\
        RVM & Rotating Vector Model \\
        SG  & Steepest Gradient \\
        $t$-mode & Transverse mode \\
        u-GMRT & upgraded Giant Metrewave Radio Telescope\\
        X-mode & Extraordinary mode \\
\end{tabular}
}

%
%

\begin{adjustwidth}{-\extralength}{0cm}

\reftitle{References}


\bibliography{References.bib}

\PublishersNote{}
\end{adjustwidth}
\end{document}